\documentstyle[12pt,aaspp4]{article}

\lefthead{Shioya et al.}
\righthead{Spectrophotometric evolution of truncated spirals}

\begin{document}
\title{Spectrophotometric evolution of spiral galaxies with truncated
star formation: 
An evolutionary link between spirals and S0s in distant clusters}

\author{Yasuhiro
 Shioya} 
\affil{Astronomical Institute, Tohoku University, Sendai, 980-8578, Japan} 

\and

\author{Kenji Bekki,  Warrick J. Couch, and Roberto De Propris}
\affil{School of Physics, University of New South Wales, 
Sydney 2052, Australia}

\begin{abstract}
A one-zone chemo-spectro-photometric model is used to investigate the
time evolution of disk galaxies whose star formation is truncated, 
and to determine the dependence of this evolution on the previous
star formation history and the truncation epoch.
Truncated spirals show red colors $\sim 1$\,Gyr after truncation, 
and evolve spectrally from an e(b) type, down through the e(a), 
a+k, and k+a classes, to finally become a k type. The exact behavior
in this phase depends on the truncation epoch and the star formation
history prior to truncation. 
For example,  earlier type disks show redder colors and do not show a+k-type spectra
after truncation of star formation.
We also discuss a possible evolutionary link
between the k-type galaxies with spiral morphology found in distant
clusters, and present-day S0s, by investigating whether truncated spirals
reproduce the infrared color-magnitude relation of Coma galaxies.
We suggest that only less luminous, later-type disk galaxies whose star formation
is truncated at intermediate and high redshifts can reproduce the red
$I-K$ colors observed for S0s in the Coma cluster.

\end{abstract}


\section{Introduction}

Unravelling the star formation histories of galaxies in rich clusters is 
one of the major issues in extra-galactic astronomy. Spectroscopy of these
galaxies has provided a number of key observations which are fundamental
to understanding star formation in the dense cluster environment: The
origin of galaxies with unusual spectra containing strong H$\delta$
absorption and no observable [OII] emission (the a+k/k+a types previously
defined as "E+A"; Dressler \&
Gunn 1983, 1992), the presence of starburst and poststarburst galaxies 
and thus evidence of a starburst evolutionary cycle (Couch \& Sharples
1987; Barger et al. 1996), the importance of a gradual decline of star
formation in explaining the observed radial gradients in the colors and
spectroscopic properties of cluster galaxies (Abraham. et al. 1996; Morris
et al 1998; Balogh et al. 1998, 1999), the connections between morphology
and spectral type (Dressler et al. 1999), and the recognition that dust
is an important ingredient to the starburst phenomenon (Poggianti et al.
1999; Smail et al. 1999). In addition, the star formation properties of 
galaxies in rich clusters are remarkably different to those in the field
(e.g., Dressler et al. 1985; Balogh et al. 1999). 

Interpreting these observations in terms of which physical processes
are responsible for the environmental differences in star formation
properties has not been straightforward. An episodic increase in star
formation rate (i.e., starburst) triggered by some cluster-related
physical process, followed by the abrupt truncation of star formation, 
have been a common interpretation of the observed spectral properties of
distant cluster galaxies (Couch \& Sharples 1987; Barger et al. 1996).
However, several authors have claimed that simple truncation or a gradual
decline of star formation can equally well explain the observations
(Abraham et al. 1996; Morris et al. 1998; Balogh et al. 1998). Even
with more recent data on the incidence and strength of H$\alpha$
emission amongst distant cluster galaxies (Couch et al. 2000;
Balogh \& Morris 2000), it is still unclear which of these two 
interpretations is most plausible.

An important aid in interpreting the spectral signatures of galaxies
and thus clearly distinguishing between these different star formation
histories, is detailed modeling of their spectro-photometric properties
within these different evolutionary frameworks. This approach has already
provided a much better understanding of starbursts going on within 
cluster galaxies, with their spectral evolution through and after
the burst well determined (Poggianti \& Barbaro 1996, Poggianti et al. 
1999), and the effects of dust and selective absorption also quantified
(Shioya \& Bekki 2000; Shioya, Bekki, \& Couch 2000). However, in
no way can this modeling effort be considered complete, with other
processes such as the simple {\it truncation} of star formation and 
the {\it chemical evolution} associated with the observed star formation
activity yet to be fully explored. 

Investigation of the truncation process is not only motivated by the
desire to understand the spectroscopic evolution that accompanies it, but
also to gather new clues as to the origin of S0s, since the formation
of this morphological class is suggested to be closely associated with 
the truncation of star formation in spirals (e.g., Larson, Tinsley, \&
Caldwell 1980). Recent morphological studies of distant cluster
populations have pointed to a strong evolution in the Sp/S0 mix with
redshift, with clusters in the past having a much smaller S0 fraction
(and a correspondingly larger Sp fraction) than their present-day
counterparts (Dressler et al. 1997; Fasano et al. 2000). The morphological
transformation of spirals into S0 galaxies would also appear to be a key
feature of cluster galaxy evolution over the last third of a Hubble time.

This demise of cluster spirals is also evidenced in a star formation sense 
through the presence of `passive spirals' (Couch et al. 1998,
Dressler et al. 1999), viz., galaxies which have a spiral morphology but
whose lack of star formation is betrayed by their passive, k-type
spectra. This would suggest that star formation ended unspectacularly
in such objects, either through truncation or just a general run down.
However, little if any modeling has been done to assess whether
this cessation of star formation activity, which must have occurred
several billion years before the epoch of observation (for the galaxy to
have evolved to a k-type), is consistent with the retention of spiral arm
structure and such systems being the progenitors of present-day S0s.

The purpose of this paper is to specifically investigate the {\it truncation}
of star formation in disk galaxies, in order to obtain a more
comprehensive understanding of the spectroscopic and photometric evolution
that results from this process. We do so in an explicitly self-consistent 
manner by using a one-zone chemical evolution model. In particular, we
investigate the evolution in both the spectral properties --  
quantified by the equivalent widths of the [OII]$\lambda$3727/H$\delta$
emission/absorption line features -- and the optical/infrared colors
following truncation, and determine its dependence on the truncation
epoch and the star formation activity prior to this event. 
Using recent observational data on the infrared color-magnitude relation
in the Coma cluster (Eisenhardt et al. 2001) as a test, we also examine
whether the truncation of spirals is a viable mode of producing 
present-day S0 galaxies. 

The layout of this paper is as follows: 
In \S 2, we summarize the numerical models used in the
present study and describe briefly the methods employed for
deriving galaxy spectral energy distributions (SEDs) and their
associated spectroscopic properties. In \S 3, we present our numerical
results for the time evolution of colors, EW([OII]), EW(H$\delta$), and
H$\alpha$ luminosities. In \S 4, we discuss an evolutionary link between
S0s and `passive' spirals formed by the truncation of star formation. 
The conclusions of the present study are given in \S 5. 

\section{Model}

We adopt a one-zone chemo-spectrophotometric  model 
of a disk galaxy with abrupt truncation of star formation  and thereby investigate 
the time evolution of UV, optical, and near-infrared colors and  
of emission and absorption lines such as H$\alpha$, [OII], and H$\delta$. 
The present model is essentially the same as that adopted in
our previous studies (Shioya \& Bekki 1998, 2000)
and the details of the treatment of chemical evolution are
already given in Shioya \& Bekki (1998). Dust effects are  not included
in the present model.
The chemical evolution of galaxies is followed by using the model 
described by  Matteucci \& Tornamb\`{e} (1987) which includes the  
metal-enrichment processes of Type Ia and II supernovae (SNIa and SNII). 
We adopt a  Salpeter initial mass function (IMF), 
$\phi(m) \propto m^{-1.35}$, with upper mass limit 
$M_{\rm up}=120M_{\odot}$ and lower mass limit 
$M_{\rm low}=0.1M_{\odot}$. 
We calculate photometric properties of galaxies as follows. 
The monochromatic flux of a galaxy having  age $T$, $F_{\lambda}(T)$ is;
\begin{equation}
F_{\lambda} (T) = \int_0^T F_{\rm SSP,\lambda}(Z,T-t) \psi(t) dt, 
\end{equation}
where $F_{\rm SSP,\lambda}(Z,T-t)$ is the monochromatic flux of 
a single stellar population with age $T-t$ and metallicity $Z$, and 
$\psi(t)$ is the time-dependent star formation rate described later. 
In the present study, we use the spectral library GISSEL96 which is the 
latest version of Bruzual \& Charlot (1993).

In this paper, we use the  infall disk model (e.g., Arimoto,
Yoshii, \& Takahara 1992) where  a disk galaxy is assumed to be gradually
built up by gas infalling from the outer halo region.
We adopt the Sa, Sb, and Sc models described in Arimoto et al. (1992) as
reasonable and realistic disk models.  
The star formation history of a disk galaxy in
the present infall model is characterized by two  epochs.  
The first is the epoch of galaxy formation 
($T_{\rm form}$ or $z_{\rm form}$) at which gaseous infall onto a disk
galaxy and the subsequent gas consumption by star
formation at a moderate rate begins. 
In the present study, ($T_{\rm form}$ ($z_{\rm form}$)  is 0 Gyr (5) 
for all models. The second is $T_{\rm trun}$ ($z_{\rm trun}$) 
at which both gaseous infall and star formation in the disk are
abruptly terminated. In the present study, $T_{\rm trun}$ ($z_{\rm trun}$)
is considered to be a free parameter. 
The gas infall rate for a disk galaxy ($A(t)$) is assumed to be 
proportional to the gas mass fraction of the reservoir gas ($f_{g{\rm
H}}$): 
\begin{equation}
A(t) = a f_{g{\rm H}},
\end{equation}
where $a$ is a parameter which controls the infall rate of a disk. 
This parameter $a$ has the values:
\begin{equation}
a = \left\{
\begin{array}{lcl}
\tau_{\rm in}^{-1} \; \; \; & & 
{\rm for} \; \; T_{\rm form}  \le T < T_{\rm trun}, \\
0   & & {\rm for} \; \; T_{\rm trun} \le T , \\
\end{array}
\right.
\end{equation}
where $\tau_{\rm in}$ gives the time scale of gas accretion.
The adopted values of  $\tau_{\rm in}$ are 
2.41 (Sa), 3.13 (Sb), and 6.76 Gyr (Sc), 
all of which are  chosen from the values
listed in Table 1 of Arimoto et al. (1992).  
We adopt three values $T_{\rm trun}$ = 4.46,
7.64, and 9.45,  each of which corresponds
to $z$ = 1.0, 0.4, and 0.2,  respectively,
for $H_{0}$ = 65 km ${\rm s}^{-1}$  ${\rm Mpc}^{-1}$
and $q_{0}$ = 0.05 (i.e., The corresponding present age
of the universe at $z$ = 0 is 13.8 Gyr).

The star formation rate is assumed to be proportional to the gas mass
fraction ($f_g$): 
\begin{equation}
\psi(t)=kf_g,
\end{equation}
where $k$ is a parameter controlling  the star formation rate. 
This parameter $k$ has values: 
\begin{equation}
k = \left\{
\begin{array}{lcl}
k_{\rm disk} \; \; \; & & {\rm for} \; \; T_{\rm form}  \le T < T_{\rm trun}, \\
k_{\rm trun}   & & {\rm for} \; \; T_{\rm trun} \le T . \\
\end{array}
\right.
\end{equation}
The value of $k_{\rm disk}$, controlling the star formation histories of
disk galaxies, is suggested to be different between different Hubble types  
(e.g., Arimoto, Yoshii, \& Takahara 1992). We adopt here $k_{\rm disk}$
values of 0.532 ${\rm Gyr}^{-1}$ (Sa), 0.409 (Sb), and 0.109 (Sc).

From now on, the above  model of truncated star formation is referred
to as the ``TF'' (truncated star formation) model.  
In order to reveal more clearly  the importance of the truncation of star 
formation in the spectral evolution of disk galaxies, 
we investigate two other models for comparative purposes: 
One is the ``TI''  (truncated infall) model in which 
only gaseous infall is assumed to be truncated at $T$ =  $T_{\rm trun}$
and star formation can continue even after  $T_{\rm trun}$. 
The  other is the ``CF''  (continuously star-forming) model in which
star formation is assumed to continue without truncation of star formation
and gas infall (i.e., exactly the same as that adopted in Arimoto et al.
1992). Accordingly, $k_{\rm trun}$ in a disk is set to be 0 for TF models
and the same as that of $k_{\rm disk}$ for CF and TI models. 
By comparing the results of these three models, we can
investigate the effects of truncating star formation
on the spectral evolution of disk galaxies. Furthermore, by changing
$T_{\rm trun}$, we can investigate how the epoch of
truncation of star formation (and gas infall)  
controls the spectral evolution of a disk galaxy.

To derive the fluxes for various gaseous emission lines 
(H$\delta$ and [OII]) in starburst galaxies, we first 
calculate the number of Lyman continuum photons, $N_{\rm Ly}$, 
by using the SED.
Here, we introduce the parameter $f_{\rm ion}$ 
which determines what fraction of Lyman continuum photons 
can be used for ionizing the surrounding gas
and we show only the results of models with $f_{\rm ion}$ =  1.0.
Previous studies assume that  
the value of $f_{\rm ion}$ is 0.2 
in Guiderdoni \& Rocca-Volmerange (1987) and 
0.7 in Fioc \& Rocca-Volmerange (1997).
Our results do not depend so strongly on $f_{\rm ion}$ if its
value is within a plausible range.
If all the Lyman continuum photons are used for ionizing the surrounding gas, 
the luminosity of H$\beta$ is calculated according to the formula: 
\begin{equation}
L({\rm H}\beta) ({\rm erg \; s^{-1}}) = 4.76 \times 10^{-13} N_{\rm Ly} ({\rm s^{-1}})
\end{equation}
(Leitherer \& Heckman 1995). 
To calculate luminosities of other emission lines, 
e.g., [OII] and H$\delta$, we use the relative luminosity 
to H$\beta$ luminosity tabulated in PEGASE 
(Fioc \& Rocca-Volmerange 1997) which is calculated 
for an electron temperature of 10000\,K and 
an electron density of 1\,cm$^{-3}$. 
We derive the strength of absorption lines based on
GISSEL96 (Bruzual \& Charlot 1993).
Thus the SED derived in the present study consists of 
stellar continuum, gaseous emission, and stellar absorption.

By using the above model,
we investigate: (1)\,how the photometric and spectroscopic properties
of disk galaxies evolve with time after the truncation of star formation,
and (2)\,whether disk galaxies, after the truncation of star formation, 
can be on the ($I-K$)-$M_{\rm K}$ and ($U-V$)-$M_{\rm V}$ color-magnitude relations 
observed in the Coma cluster. 
In the first investigation, we analyze in particular the time evolution
of the $U-V$, $B-V$, and $I-K$ colors and the spectroscopic properties on
the EW([OII])-EW(H$\delta$) plane.
In order to show more clearly the present numerical results, 
we first show the results of a set of fiducial Sb disk models 
($T_{\rm trun}$ = 7.64 Gyr, $k_{\rm disk}$ = 0.409 Gyr$^{-1}$).
In this set of models, we show the typical 
evolution of disks with truncated star formation
by comparing the three models, CF, TF, and TI.
We then show the dependences of the evolution
on the epoch of truncation ($T_{\rm trun}$ or $z_{\rm trun}$)
and on the initial star formation histories of the disks ($k_{\rm disk}$).
In total, we describe the results of 21 models with different
$T_{\rm trun}$ and $k_{\rm disk}$ values to 
elucidate the importance of each  parameter.
For convenience, the Sb model with truncation
of gas infall at $T_{\rm trun}$ = 7.64 Gyr is represented by 
${\rm Sb}_{\rm TI}^{7.64}$:
We investigate ${\rm Sa}_{\rm TI}^{4.46}$, ${\rm Sa}_{\rm TI}^{7.64}$, ${\rm Sa}_{\rm TI}^{9.45}$, 
${\rm Sa}_{\rm TF}^{4.46}$, ${\rm Sa}_{\rm TF}^{7.64}$, ${\rm Sa}_{\rm TF}^{9.45}$, 
${\rm Sa}_{\rm CF}$, 
${\rm Sb}_{\rm TI}^{4.46}$, ${\rm Sb}_{\rm TI}^{7.64}$, ${\rm Sb}_{\rm TI}^{9.45}$, 
${\rm Sb}_{\rm TF}^{4.46}$, ${\rm Sb}_{\rm TF}^{7.64}$, ${\rm Sb}_{\rm TF}^{9.45}$, 
${\rm Sb}_{\rm CF}$, 
${\rm Sc}_{\rm TI}^{4.46}$, ${\rm Sc}_{\rm TI}^{7.64}$, ${\rm Sb}_{\rm TI}^{9.45}$, 
${\rm Sc}_{\rm TF}^{4.46}$, ${\rm Sc}_{\rm TF}^{7.64}$, ${\rm Sc}_{\rm TF}^{9.45}$, 
and ${\rm Sc}_{\rm CF}$. 

In the second investigation, we analyze the photometric evolution
of disks with variously different truncation epochs and star formation
histories on the ($I-K$)-$M_{\rm K}$ color-magnitude (CM) relation observed
by Eisenhardt et al. (2001) and de Grijs \& Peletier (1999)
and ($U-V$)-$M_{\rm V}$ CM one observed by Bower et al. (1992).
Since the observed colors for S0s are composed of both disk and  
bulge contributions, we also consider the color evolution of the bulge
component on its own in this second investigation.
In addition to the disk colors calculated using the above disk models, 
we derive bulge colors and then total colors for disk galaxies
(with and without truncation) based on the bulge model described below.
We adopt almost the same bulge model as that of Arimoto \& Jablonka
(1991) in which 0.7 Gyr after the monolithic collapse of the bulge
the galactic wind blows away the remaining gas to completely stop 
star formation. In the present bulge model, $k_{\rm disk}$ = 10.0\,${\rm
Gyr}^{-1}$ and $\tau_{\rm in}=$0.125\,Gyr. We investigate the importance
of bulge colors in reproducing the E/S0 $I-K$ and $U-V$ CM relation, by 
considering a series of truncated and non-truncated disk galaxies
with different bulge mass fractions ($f_{\rm bul}$; the fraction
of bulge mass to total mass of a disk galaxy).
The values of $f_{\rm bul}$ investigated in the present study
are 0.1, 0.3, and 0.5. 
As part of this second investigation, 
we derive the $z=0$ (final)  mass needed 
to place its disk component on the ($I-K$)-$M_{\rm K}$ CM-relation 
observed for disk galaxies by de Grijs \& Peletier (1999). The  
total stellar masses adopted for the Sa, Sb, and Sc disks in the CF models
are 3.27 $\times$ $10^{10}$ $M_{\odot}$, 1.85  $\times$ $10^{10}$ $M_{\odot}$,
and 4.49 $\times$ $10^{9}$ $M_{\odot}$, respectively. 
Thus, the  adopted model assumtion that 
a later type disk (e.g., S(c)) is less lumninous and less massive
and has bluer colors are consistent both with previous results 
on the dependence of disks' colors on galactic morphology 
(e.g., Arimoto et al. 1992)
and with the observed CM relation by de Grijs \& Peletier (1999). 

Overall, we investigate the photometric evolution of 54 different models
on the ($I-K$)--$M_{\rm K}$ and ($U-V$)-$M_{\rm V}$ planes and thereby study 
under  which physical conditions (e.g., star formation histories
and truncation epoch) truncated disk galaxies can reproduce
the observed CM relations of E/S0s.
For convenience, a set of Sa (Sb and Sc) models
with variously different truncation epochs, bulge mass fraction,
and truncation types (CF, TF and TI)
are referred to as model sequence S1 (S2 and S3, respectively).
These model sequences are summarized in Table 1.
We stress that the derived EW([OII]) in the present calculation  
is appreciably larger than
the typical value of the observed EW([OII]).
This is because we use the emission line ratio of [OII]/H$\delta$
(3.01 throughout the present paper) listed in the table of the PEGASE
(Fioc \& Rocca-Volmerange 1997) in order to calculate [OII] emission
line strength of a truncated spiral. 
Shioya, Bekki, \& Couch (2001) have already demonstrated 
that the [OII] emission
line strength calculated by the PEGASE is appreciably (a factor of $\sim$ 1.5)
larger than that by the photoionization code CLOUDY (Ferland et al. 1998). 
Therefore the present results should be more carefully interpreted
especially when we determine whether a spiral galaxy with star formation
shows the spectral types e(b) or e(c). 
More discussions on  advantages and disadvantages in 
using the adopted PEGASE code are  given in Shioya et al. (2001).

\placefigure{fig-1}
\placefigure{fig-2}
\placefigure{fig-3}
\placefigure{fig-4}
\placefigure{fig-5}

\section{Results}

We show first the results for our fiducial models 
(${\rm Sb}_{\rm TF}^{7.64}$, ${\rm Sb}_{\rm TI}^{7.64}$, and ${\rm Sb}_{\rm CF}$)
and then describe 
the parameter dependences of the photometric and spectroscopic evolution
of disk galaxies. The star formation histories of the models with and
without truncation (of star formation or gas infall) are given in Figures
1 and 2. Mean star formation rate is higher for earlier type disks
than later ones,
and consequently gas mass fraction is more rapidly decreased for earlier
type disks than later type ones (Figure 1).  
As is shown in Figure 2, 
the star formation rate in a TI model becomes lower
than that  of the  CF model with the same parameters as those in the TI one 
soon after truncation of gas infall 
whereas the star formation rate in a TF model becomes abruptly 0
after truncation of star formation.

\subsection{Fiducial model}

As is shown in Figure 3, all of the $U-V$, $B-V$, and $I-K$ colors for 
${\rm Sb}_{\rm TF}^{7.64}$  become rapidly
redder after truncation in comparison to those of ${\rm Sb}_{\rm CF}$. 
This is due essentially to the aging of the stellar populations after
the truncation of star formation in ${\rm Sb}_{\rm TF}^{7.64}$ . 
The difference in colors between the two models 
(${\rm Sb}_{\rm TI}^{7.64}$ and ${\rm Sb}_{\rm CF}$)  
depends on wavelength with it being more clearly observed in the UV and
optical bands ($U-V$ and $B-V$) than in $I-K$. This reflects the fact
that the contribution from the young stellar populations
to color evolution is much more dominant at UV and optical wavelengths
than in the near-infrared. The color difference 
between ${\rm Sb}_{\rm TF}^{7.64}$  and  ${\rm Sb}_{\rm CF}$ 
after 1\,Gyr (corresponding roughly to the epoch when the difference
reaches a maximum) is estimated to be 0.7\,mag for $U-V$, 0.3\,mag for
$B-V$, and 0.05\,mag for $I-K$.

${\rm Sb}_{\rm TI}^{7.64}$, on the other hand, does not show such a drastic color
difference with respect to ${\rm Sb}_{\rm CF}$, simply because star formation 
still continues after the truncation of gas infall. The color difference
between ${\rm Sb}_{\rm TI}^{7.64}$ and ${\rm Sb}_{\rm CF}$ is smallest at longer wavelengths.
What is interesting here is that although the UV and optical colors
in  ${\rm Sb}_{\rm TF}^{7.64}$
 are appreciably redder than those of  ${\rm Sb}_{\rm TI}^{7.64}$   model after
truncation, the $I-K$ color is redder 
in ${\rm Sb}_{\rm TI}^{7.64}$  than in ${\rm Sb}_{\rm TF}^{7.64}$ 
model. This is essentially because chemical evolution proceeds in ${\rm Sb}_{\rm TI}^{7.64}$ 
model even after truncation, with more metal-rich components being formed
that make the $I-K$ colors redder in comparison to ${\rm Sb}_{\rm TF}^{7.64}$. The
UV and optical colors, however, are more strongly affected by the young
stellar populations than metallicity, and thus are bluer with respect
to ${\rm Sb}_{\rm TF}^{7.64}$.

Spectroscopic evolution after truncation is more drastic in ${\rm Sb}_{\rm TF}^{7.64}$ 
than in ${\rm Sb}_{\rm TI}^{7.64}$, as is shown in Figure 4.
For example, EW(H$\delta$) reaches  $\sim$ 7 $\rm\,\AA$ in ${\rm Sb}_{\rm TF}^{7.64}$ 
whereas it reaches only  $\sim$ 3  $ \rm \AA$  in ${\rm Sb}_{\rm TI}^{7.64}$.
H$\alpha$ luminosity and EW([OII]) gradually decline after truncation
of gas infall in the TI model and consequently  
the maximum differences between ${\rm Sb}_{\rm CF}$ and ${\rm Sb}_{\rm TI}^{7.64}$ 
and ${\rm Sb}_{\rm TI}^{7.64}$  are $\sim$ 10 $\rm \AA$ for EW([OII]) and 
1.5 $\times$ $10^{34}$ W for H$\alpha$ luminosity. 
These results confirm that ${\rm Sb}_{\rm TI}^{7.64}$  shows a gradual decline in
emission line strength without strong H$\delta$ absorption.
Figure 5 describes the evolution of the three models 
on the EW([OII])-EW(H$\delta$) plane
which was originally introduced by Dressler et al. (1999) and 
considered to be a diagnostic of the star formation activity.
${\rm Sb}_{\rm TF}^{7.64}$ evolves from e(b), to e(a), to a+k, to k+a, and finally to k
whereas ${\rm Sb}_{\rm TI}^{7.64}$  only exhibits an e(b) or e(c) type spectrum. 
${\rm Sb}_{\rm TF}^{7.64}$  provides an evolutionary link between the different spectral
types observed in clusters.  

What should also be remarked here is that the time spent by ${\rm Sb}_{\rm TF}^{7.64}$ 
in the e(a) region is shorter than $10^7$\,yr. 
This is in striking contrast to the dusty starburst models 
(e.g., Shioya \& Bekki 2000) where the time spent within the e(a) region
is rather long ($\sim$ 0.7 Gyr). These results imply that
the e(a) galaxies observed in distant clusters (e.g., Poggianti et al. 1999)
are more likely to be dusty starburst galaxies 
rather than spirals with truncation of star formation. 
Accordingly  our numerical model demonstrates that a spiral galaxy whose
star formation is truncated, not only evolves into a 
a+k/k+a galaxy but also eventually becomes a passive
spiral with a k-type spectrum.
Our numerical results thus confirmed the evolutionary path 
from truncated spirals into k-type one, which has been suggested
by Poggianti et al. (2000). 

\placefigure{fig-6}
\placefigure{fig-7}
\placefigure{fig-8}
\placefigure{fig-9}
\placefigure{fig-10}
\placefigure{fig-11}

\subsection{Parameter dependence}

Figure 6 describes how the truncation epoch (of star formation and 
gas infall), represented by $T_{\rm trun}$ or $z_{\rm trun}$, 
controls the color evolution of the Sb disk models
(${\rm Sb}_{\rm TF}^{4.46}$, ${\rm Sb}_{\rm TF}^{7.64}$, and ${\rm Sb}_{\rm TF}^{9.45}$). 
After
truncation, the $U-V$,  $B-V$, and $I-K$  colors are 
discernibly redder for the model with  later truncation (smaller $z_{\rm trun}$) 
in the TF models, essentially because disks 
with later truncation have more metal-rich stellar components due to
chemical evolution. This result implies that if S0s are formed by
the truncation of star formation, then the earlier the epoch the
bluer their colors will be.
The three colors in the TI models, on the other hand, are redder for the model 
with  earlier truncation (larger $z_{\rm trun}$).
This is because a disk galaxy whose gas infall occurs at a later time,
will have a larger number of young blue stellar components. 

Irrespective of $T_{\rm trun}$ ($z_{\rm trun}$), the $U-V$ and $B-V$ colors are redder 
in the TF than in the TI model, whereas the reverse applies for the $I-K$
color. As is shown in Figure 7, the above results on the $T_{\rm trun}$
dependences of color evolution for the Sb disk models are basically true
for the Sc models. There is only one difference in the parameter
dependence between the Sb and Sc models, and that is for the Sc model, 
the $I-K$ color for $T_{\rm trun}$=9.45  ($z_{\rm trun}$=0.2) is appreciably redder for the TF
case than for TI. This reflects the fact that in the Sc model, where the
star formation is rather slow, gas still remains in the disk after
truncation thus prolonging the star formation and causing younger stars
to be formed. As is shown in Figure 8, there are no remarkable
differences in the $T_{\rm trun}$ dependences in color evolution between
the Sb and Sa models.

Figure 9 indicates two important $T_{\rm trun}$ dependences of spectral
evolution on  the EW([OII])-EW(H$\delta$) plane in the Sb disk models.
One is that in the TF model with earlier truncation (larger $z_{\rm trun}$),
the galaxy passes through the e(a), a+k, and k+a regions further to the
right, as a result of the larger EW(H$\delta$). 
This is principally because the number ratio of young stars (responsible
for the emission and absorption lines) to older ones (responsible for
the continuum),
which is an important determinant of EW(H$\delta$), is larger for
the model with the earlier truncation (larger $z_{\rm trun}$).    
The other is that the TF model with the later truncation (smaller $z_{\rm trun}$) 
does not show an a+k spectrum.  
The evolution of the TI models does not depend so strongly on $T_{\rm trun}$:
irrespectively of $T_{\rm trun}$, the galaxies show either an e(b)- or
e(c)-type spectrum throughout their entire evolution.

As is shown in Figure 10, there are two remarkable differences in the
$T_{\rm trun}$ dependences between the Sb and the Sc models. 
One notable difference for the Sc models is that they all go through an
a+k phase after truncation, no matter what the value of $T_{\rm trun}$ is.
The other is that the time-scale for the galaxy to show a k-type spectrum 
after truncation, is rather long for the different $T_{\rm trun}$ models. 
(Actually, the Sc models do not show a k-type spectrum at the present
epoch, $z=0$.) This implies that k-type `passive' spirals formed from
late-type Sc disks are less likely to be observed in distant clusters.
The $T_{\rm trun}$ dependences derived for the Sa models are essentially
the same as those for the Sb models, as is shown in Figure 11.

\placefigure{fig-12}

\subsection{Comparison  with the color-magnitude relation of Coma cluster}

Figure 12  shows whether TF models for disk galaxies with Sc-type star
formation can explain the ($I-K$)-$M_{\rm K}$ color-magnitude
(CM) relation observed for S0s in the Coma cluster. Here, the initial gas
mass of the disk in each model is chosen such that the color and the
magnitude of the disk at $z=0$ is on the ($I-K$)-$M_{\rm K}$ CM relation 
observed for disk galaxies by de Grijs \& Peletier (1999). Therefore an Sc
model with a lower final metallicity (and hence bluer colors) is less
luminous compared to the early-type spiral models (Sa and Sb). Figure 12
demonstrates the following four important results with regard to the 
transformation of spirals into cluster S0s through the truncation of their
star formation: First, TF models  at $z=0$ show red enough $I-K$ colors
to reproduce the CM relation, which implies that truncated Sc disks at
intermediate and high $z$ can become present-day S0s. Second, an Sc disk
whose star formation is truncated at a later time, has a brighter $M_{\rm
K}$ and bluer $I-K$ color. Third, at a given truncation epoch, models 
with a larger initial bulge show a brighter $M_{\rm K}$ and redder $I-K$
color after truncation. These results imply that if star formation in Sc
disks is completely truncated, they become S0s that are on the observed
($I-K$)-$M_{\rm K}$ CM relation at $z=0$, though the positions of the
S0s in the CM relation depends strongly on the epoch of truncation and 
initial bulge mass.

What should be emphasized here is that although the present Sc models
with truncation of star formation can explain the observed CM relation of S0s,
they do not show k-type spectra that are typical for S0s in clusters.
The reason for the model's failure to reproduce the k type
is due to the fact that in the adopted population synthesis model GISSEL 96,
H$\delta$ can not be so small as to show the k type if the metallicity
of the adopted chemical evolution model  (i.e., corresponding to
the present Sc model) is small.
The aftermath of the Sc model with truncation of star formation 
is a less luminous dwarf S0. 
Therefore, whether or not the Sc model with truncation of star formation
can be rejected as a S0 model
depends on the observed spectral type of less luminous dwarf S0s with $M_{v}$ $>$ -19 mag
in low and intermediate redshifts clusters of galaxies.
However,  owing to the lack of observations on the spectral types of dwarf S0s 
in clusters environments,
it is currently impossible for us to discuss this issue.
Thus we suggest  that future very deep spectroscopic
observations of dwarf S0s in clusters will provide the answer as to
whether the Sc model with truncation of star formation
is plausible and realistic for S0 formation.

Figure 13 shows whether TI models for disk galaxies with Sc-type star
formation can explain the observed $I-K$ color-magnitude (CM) relation 
for Coma S0s. TI models at $z$ = 0 have appreciably bluer colors than 
their TF counterparts, and are less likely to be on the Coma
($I-K$)-$M_{\rm K}$ CM relation (in particular, for models with smaller
bulges or later truncation epochs). This is essentially because ongoing
star formation in the TI models with Sc-type star formation results in
bluer $I-K$ colors. The tendency for the TI models to show appreciably
bluer colors after truncation (cf. the TF models) can be seen in the other
models with Sa-type and Sb-type star formation. These results imply that
simple truncation of gas infall alone, cannot explain the observed red
colors of S0s and thus after the truncation of gas infall, star
formation also needs to be truncated by some mechanism in order to
reproduce the CM relation.  

Compared to the Sc TF models, the Sb TF models show smaller differences
in their $I-K$ colors at $z=0$ when comparing those with and without
truncation (Figure 14). Furthermore, the truncated Sb models at $z=0$
have appreciably bluer (typically 0.1\,mag) colors than those observed, 
implying that Sb disk galaxies are less likely to become S0s after
truncation of gas infall or star formation than are Sc galaxies.
However, considering the observed scatter in the $I-K$ colors of both disk
and S0 galaxies, it can be seen from Figure 14 that some fraction of the
observed S0s (in particular, those with bluer colors) could plausibly 
be formed from truncated Sb galaxies. As is shown in Figure 15, the
parameter dependences of the TI models with Sb-type star formation
histories are essentially the same as those derived for the Sc models.

The dependences of the colors and magnitudes of truncated disks on
the initial bulge mass fraction ($f_{\rm bul}$) and truncation epochs
in the Sb models, is essentially the same as those of the Sc models.
Figures 16 and 17 show the evolution of truncated disks on the
($I-K$)-$M_{\rm K}$ CM relation for the TF and TI models, respectively,  
for Sa-type star formation and with different truncation epochs and
$f_{\rm bul}$. These figures confirm that earlier-type disk galaxies (Sa
and Sb) show bluer colors after truncation than those observed and
thus are less likely to be the progenitors of S0s than later-type disk
galaxies (Sc). The parameter dependences of the evolution of Sa disks on
the $I-K$ CM relation are essentially the same as those derived for
the Sb and Sc disk models.

The parameter ($T_{\rm trun}$, $f_{\rm bul}$, and star formation
type) dependences  of photometric evolution of truncated spirals
in the ($U-V$)-$M_{\rm V}$ plane are essentially the same as those
derive in ($I-K$)-$M_{\rm K}$ plane (See right panels in Figure 12 -- 17)
and summarized as follows.
Firstly only TF models with Sc-type star formation
can show enough red colors to be close to the observed S0 CM relation:
truncated Sb and Sa disks can not be on the S0 ($U-V$)-$M_{\rm V}$ CM relation.
This result implies that since spirals with Sc-type star  formation
corresponds to low luminous ones in the present study,
only low luminous spirals can become S0s after truncation of star formation.
Secondly, irrespectively of disk's star formation history and bulge mass fraction,
spirals with earlier truncation (larger $z_{\rm trun}$) are more likely to
show redder $U-V$ color and become  fainter (i.e., larger $M_{\rm V}$)
after truncation of star formation.
Thirdly, TF models are more likely to be redder than TI ones for each
star formation type.


It follows from Figures 12, 14, and 16 that it is the less luminous disk
galaxies that are more likely to reproduce the observed S0 $I-K$ colors at
$z=0$, after having their star formation truncated. Consequently, the 
origin of more luminous S0s might not be so closely associated with
the truncation of star formation. The basic reason for this is that 
color evolution after truncation is rather less dramatic for more massive
disk galaxies because of the larger fraction of (older) stars that have 
already been formed before truncation in these galaxies.  
There are mainly two interpretations for the $apparent$ failure 
of truncated star formation models (TF) to reproduce the red $I-K$ colors 
observed for the more luminous S0s (thus CM relation).

First, this result simply means that truncation of star formation is not 
a major formation mechanism for the more luminous S0s in clusters.
Shioya \& Bekki (1999) have already suggested that massive secondary 
starbursts and the resultant chemical enrichment are important
requirements for explaining the redder S0 colors, based on one-zone models
similar to those adopted in the present study. In this first interpretation,
more luminous S0s are formed not by the simple truncation
of star formation, but by some physical mechanism which involves
significant chemical enrichment, hence giving S0s their redder colors.

The second interpretation is that the present study  does  not correctly
model the progenitor disk galaxies of more massive cluster S0s, so that
the rather red S0 colors that are observed are not reproduced by the 
truncation models. It is possible that the progenitor disk galaxies of
more luminous S0s are systematically redder than disk galaxies on the
present-day CM relations: the S0 progenitor galaxies are intrinsically
different to disk galaxies observed at $z=0$. In this interpretation, if
the S0 progenitor galaxies are appreciably redder than the present-day
ones, then the S0 CM relation can still be explained by simple truncation.

It is, however, not at all clear why the disk galaxy progenitors of more
massive cluster S0s would be intrinsically redder. One possible reason is
that the slope of their initial mass function is not as steep as that of
present-day disk galaxies. It has already been demonstrated by Arimoto \&
Jablonka (1991) and Kodama \& Arimoto (1997) that the IMF of more massive
early-type galaxies cannot be as steep (the exponent $\alpha$ $\sim$
$-1.1$) as the Salpeter IMF ($\alpha$ = $-1.35$), if the rather red colors
of early-type galaxies are to be reproduced. However, the question arises 
as to why the IMF of more massive S0 progenitor disks is so different
from that of present-day disk galaxies? It is not clear at all what the
main physical mechanism is for the speculated difference in the IMF slope,
which leads us to suggest that this second interpretation is not so
plausible. 

An important implication of the first (and perhaps more plausible)
interpretation is that the population of blue `Butcher-Oemler' galaxies 
(Butcher \& Oemler 1978) in
intermediate redshift clusters that are the putative progenitors of 
present-day S0s (Dressler et al. 1997), should comprise mainly under-luminous, 
late-type ($\sim$Sc) disk galaxies. As $HST$-based high-resolution imaging
studies of distant clusters have shown, this is generally the case (Couch et al.
1994, Dressler et al. 1994, Smail et al.1997).
Furthermore recent photometric studies (optical and near-infrared colors)
and spectroscopic ones (H$\beta$) for S0s in intermediate clusters of galaxies
have also suggested that fainter S0s appear to have relatively young
(2--5 Gyr) stellar populations (Kuntschner \& Davies 1998; Smail et al. 1998, 2001).
Accordingly an important question is {\it whether 
or not only less luminous S0s show clear redshift evolution of their number 
fraction. }
Since  the number evolution of S0s is estimated for $all$ S0s (with variously
different luminosities), this question is not observationally answered. 
If the number evolution is observationally confirmed to be 
remarkably seen $only$ in less luminous S0s,
the truncation of star formation can be one of plausible  mechanisms of less luminous S0 formation.
Thus we stress that future observational studies on
the dependences of S0 number evolution of S0 luminosities
are doubtlessly worthwhile for determining the relative importance
of the truncation of star formation in S0 formation.
We lastly suggest that the observational result (Gregg 1989; Bothun \& Gregg 1990) that 
disks have younger ages than bulges in S0s can reflect the fact that
low luminous S0s are formed from low luminous spirals 
(Sc-type) with higher recent star formation 
and thus with younger mean ages of disks.


\section{Discussion}

\subsection{An evolutionary link between passive spirals and S0s}

Gaseous dissipation in a spiral galaxy and the subsequent formation of stars
with small random motions, are suggested to suppress the increase of velocity 
dispersion (thus the increase of Toomre's $Q$ parameter: Toomre 1964) and
consequently maintain its spiral structure (Sellwood \& Carlberg 1984).
Conversely, if gas is removed very efficiently by some physical processes (e.g.,
ram-pressure stripping) or consumed rapidly by star formation, the spiral arms 
will disappear after several disk rotation periods ($\sim 10^{9}$\,yrs; Sellwood
\& Carlberg 1984).
 As is demonstrated by the present study, a  disk 2--3 Gyr after the
removal of gas (or truncation of star formation)
can be observed as a galaxy  with a k spectrum. 
Therefore the present sepctrophotometric study and the previous
dynamical ones (on the time scale of spiral disappearance)
imply that it is very hard to observe {\it a galaxy both with clear
spiral arms and with a k spectrum}: The time scale of spiral
arms' disappearance is appreciably shorter than the evolutionary
time scale from e(b), to a+k (k+a), and to k (i.e,
we can observe  $only$ a+k/k+a spirals). 
However, we here stress that previous numerical simulations
with the total particle number of $\sim$ $10^4$ inevitably suffered
from numerical heating and thus  underestimated  the time scale
of spiral arms' disappearance.
Accordingly it is possible that the real time scale of spiral
arms' disappearance is similar to or larger than 2--3 Gyr.  
Future very  high-resolution simulations (of dynamical
evolution of isolated spiral galaxies)
with the particle
number of $10^7$--$10^8$ will predict more precisely  the time scale
of spiral arms' disappearance.

Although we can not so strongly suggest that a truncated disk
can reproduce both morphological and spectroscopic properties
of the observed  passive k-type spirals, 
we can say that an  observed k spirals
is  transformed into a S0 with featureless disk and an k spectrum
after the spiral arms disappear.
Such S0s as those formed from  k-type spirals  can
show both rather red colors and large bulge-to-disk ratios owing to disk
fading (Larson et al. 1980). Thick disk components that S0s are observed
to have (Burstein 1979) can be formed by tidal heating of cluster tidal
field and by weak galaxy interaction after the formation of S0s. Therefore, 
the truncation scenario can provide an evolutionary link between passive
spirals observed in distant clusters and S0s with the number fraction
observed to decrease with the increase of redshift.

There are, however, several problems with this truncation scenario for S0
formation: Firstly, it is not clear how the truncation scenario can
explain why the sizes of cluster S0 bulges are larger, absolutely, 
than those of late-type spirals (Dressler 1980). Since nuclear starbursts, 
followed by the growth of bulges, do not occur at all in the truncation
scenario, the absolute bulge sizes of developed S0s should nearly be
the same as those of late-type spirals. Hence the observations would
imply that either the bulges of the disk galaxies from which the SOs are
formed are larger than those of typical late-type spirals, or simple
truncation cannot explain the origin of S0s with big bulges.

Secondly, it is also unclear whether the truncation scenario can
explain the observed stellar rotation curves, velocity dispersion profiles,
and line-of-sight velocity distributions of S0s recently derived by
Fisher (1997). If these kinematical properties of bulges in late-type
disks are largely observed to be different from those of S0s, the
truncation scenario has a serious problem in explaining the kinematical
properties of S0s. More detailed comparisons of the kinematics of
the bulges in late-type disk galaxies with those of S0s are needed to
assess the validity of the truncation scenario.  

Thirdly, the truncation scenario has not yet given a clear explanation for
the difference in the observed color gradients between cluster S0s 
(e.g., Eisenhardt et al. 2001) and late-type disks (e.g., de Jong 1996).
Evolution of color gradients due to the aging of the stellar population
is rather moderate, and accordingly the difference between the gradients
in S0s and in late-type disks at low redshift can provide strong
constraints on the epoch of truncation. If future observational studies
reveal that late-type disks at intermediate redshift already have
color gradients, we can confirm whether the difference in color gradients
between present-day S0s and late-type disk galaxies at intermediate
redshift can be explained by the aging of their stellar populations after
truncation and thereby clarify  the relative importance of the truncation
scenario of S0 formation in clusters. 

\subsection{Physical mechanism of truncation of star formation}

If some fraction of S0s are actually formed by the truncation of star
formation, what physical process is responsible for the truncation ? 
Two possible mechanisms have previously been proposed: ram pressure
stripping (Gunn \& Gott 1972) and abrupt truncation of gas infall onto
disks from the halo regions (Larson et al. 1980). Although previous
numerical simulations already demonstrated that ram pressure
stripping can efficiently remove the gas within a disk (e.g., Farouki \&  
Shapiro 1980; Abadi, Moore, \& Bower 1999), recent high-resolution
hydrodynamical simulations, including turbulent and viscous stripping,
have demonstrated that this process is capable of $completely$  removing all
the gas (Quilis, Moore, \& Bower 2000).

However, these studies only showed that HI gas, which is not directly responsible
for star formation, is removed from disk galaxies. Since star formation
occurs in the cores of molecular clouds where significant amounts of
hydrogen exist in molecular (H$_{2}$) form , the complete 
removal of HI gas does not necessarily mean the complete truncation of
star formation -- star formation could continue even after the ram
pressure stripping of HI gas. The H$_{2}$-to-HI gas mass ratio is
observed to be widely different between disk galaxies, ranging from
0.2$\pm 0.1$  to 4.0$\pm$ 1.9 (Young \& Scoville 1991).
Hence the extent to which ram-pressure stripping truncated star formation
will accordingly vary.

Larson et al. (1980), on the other hand, suggested that if gas
replenishment of the disk (due to infall of halo gas) is truncated by
some mechanism,  star formation can be truncated after the rapid
consumption of the remaining disk gas.  Bekki, Couch, \& Shioya
(2001) have demonstrated that such truncation can be caused by the tidal
stripping of these gas reservoirs by cluster global tidal fields.
However, as is demonstrated by this study, such truncation of gas infall 
will not stop star formation completely, as long as the Schmidt law for
star formation is applicable. Spirals with k-type spectra cannot be
formed without other mechanisms of truncation of star formation.

Thus, it remains still unclear what physical processes are closely
associated with the truncation of star formation of late-type spirals in
cluster environments. Probably, the combination of the above two
mechanisms plus other unknown physical mechanisms all play some role
in the truncation of star formation. We note here the importance of 
the ``threshold  effect'' for star formation. Kennicutt (1989) showed that 
the onset of star formation is related to the local gravitational
instability criterion represented by the so-called $Q$ parameter and
furthermore suggested that if the $Q$ parameter for a disk
region is larger than some critical value (e.g., 1.0-1.5), star formation
does not occur in that region. Accordingly, even if a small amount of gas
remains in a spiral, star formation will not occur if, for whatever
reason, this threshold is exceeded.

Dynamical heating of the stellar components by the spiral arms in a  disk
after truncation of gas infall can increase 
the velocity dispersion of the disk and consequently lead to larger $Q$
values ($>2$; Sellwood \& Carlberg 1984; Bekki, Shioya, \& Couch 2001).
Also, weak tidal interactions with other cluster members and the 
cluster tidal field could cause dynamical heating within the disk and thus
larger $Q$ values. Accordingly, the rapid increase of the $Q$ parameter
due to dynamical heating related to the cluster environment could play
a major role in greatly suppressing star formation after the removal
of disk HI gas or the truncation of gas replenishment. If this scenario is
true for most of the passive spirals with k-type spectra (i.e.,
with no detectable star formation), such spirals can still have a small
amount of gas. Thus it will be important to investigate observationally
whether such spirals have a significant HI and/or H$_{2}$ gas content
and show large $Q$ values in their disks.

\section{Conclusions}

We have investigated, numerically, the spectroscopic (EW([OII]),
EW(H$\delta$), and $L$(H$\alpha$)) and photometric properties ($U-V$, $B-V$, and
$I-K$ colors) of disk galaxies whose star formation is truncated. Our main
conclusions are as follows:

(1)\,Late-type disks, with Sb- and Sc-type star formation histories,
undergo the following evolution in spectral type after truncation of
their star
formation: e(b)$\rightarrow$e(a)$\rightarrow$a+k$\rightarrow$k+a$\rightarrow$k.
Early-type disks with Sa-type star formation do not show a+k spectra after
truncation. These results imply that passive spirals with k-type spectra
can be formed from both late-type and early-type disk galaxies;
disks with a+k spectra are more likely to be formed from later types.
These  results thus confirmed the evolutionary path 
from truncated spirals into k-type one, which has been suggested
by Poggianti et al. (2000). 
Irrespective of their initial Hubble types, the truncated spirals show
rather red colors $\sim$ 1\,Gyr after truncation.

(2)\,Abrupt truncation of gas infall onto a disk from its gaseous halo and   
the resultant gradual decline of star formation, on the other hand, cannot
be responsible for the a+k/k+a and k-type spectra observed in clusters.
As a result of the star formation that continues after truncation, the
colors are appreciably bluer than those of models with truncation of star
formation. Also TI models show emission lines even after truncation of
gas infall.

(3)\,The spectroscopic and photometric properties after truncation of star
formation depend on the epoch of truncation such that disks which are 
truncated at later times, have redder colors and do not show a+k-type 
spectra after truncation. These properties also depend on the star
formation history of the disk before truncation such that earlier type
disks show redder colors and do not show a+k-type spectra.

(4)\,Only less luminous, later-type disk galaxies whose star formation
is truncated at intermediate and high redshifts can reproduce the red 
$I-K$ colors observed for S0s in the Coma cluster. This means that either
the truncation scenario for S0 formation is only viable for less luminous
S0s or that S0 progenitor disk galaxies are intrinsically different
from the present-day typical disk galaxies. 

(5)\,More detailed observational analysis of the nature of spiral galaxies
with k- and k+a-type spectra, such as the $K$--band luminosity function,
spatially resolved color distributions, and the HI and H$_{2}$ gas content
will be indispensable for better understanding the evolutionary links
between spiral galaxies with truncated star formation and the present-day
S0 galaxies.

\acknowledgments

We are  grateful to the referee Ian Smail for valuable comments,
which contribute to improve the present paper.
Y.S. thanks the Japan Society for Promotion of Science (JSPS) 
Research Fellowships for Young Scientists. K.B. and W.J.C acknowledge the
financial support of the Australian Research Council.

\newpage
\appendix

\section{Comparison with other observational results}
Our main purpose of the present study is not
to investigate H$\beta$ line strength and global colors of E/S0s 
observed in the present-day clusters.
However,  it might well be important
to confirm whether or not the adopted numerical model
can reproduce these observed properties of early-type
galaxies, because this investigation can  
not only demonstrate the ability of our model's reproducing
photometric and spectroscopic properties
of early-type galaxies (not discussed in the section 3) 
but also enhance the credibility of our  results.

\subsection{H$\beta$ strength}
Figure 18 describes the time evolution of TI and TF  models (no bulges) 
with three different star formation histories  (i.e., Sa-, Sb-, and Sc-type
star formation) and with different truncation epochs
on H$\beta$--$M_{\rm V}$ plane.
This figure demonstrates that the TF models  can reproduce reasonably well
the observed H$\beta$ strength ($\sim$ 2 \AA) by J{\o}rgensen (1999) well after
the truncation of star formation. The TI models, on the other hand,
can not be located in the regions where most of E/S0a are distributed
(in the H$\beta$--$M_{\rm V}$ plane). 
These results confirm that our TF model results are consistent with
the observed H$\beta$ strength at $z$ = 0 and thus strengthen  the scenario
that some S0s (in particular lower luminous S0s) can be formed from
spirals with truncation of star formation.

\subsection{Global colors  of elliptical galaxies}

By assuming early truncation  of star formation (i.e., very small $T_{\rm trun}$)
and adopting very short infall and star formation time scales (i.e.,
very small ${\tau}_{\rm in}$ and very large $k_{\rm disk}$),
we can construct an photometric evolution model of elliptical galaxies
(e.g., Kodama \& Arimoto 1997, hereafter referred to as KA).
We here adopt more metal-rich models with the slope of IMF equal to 1.10,
${\tau}_{\rm in}$ = $k_{\rm disk}$ = 0.1 Gyr, and $T_{\rm trun}$ = 0.3, 0.5, 0.8,
and 1.0 Gyr for elliptical galaxies (with galactic wind).
In these four models, earlier truncation models correspond to less massive 
elliptical ones   (KA).
We also adopt both SSPs  in  GISSEL96 and those in KA
in order to confirm whether or not our models can reproduce the observed
colors of E/S0s at $z$  = 0,  irrespectively of the adopt stellar population models.
Figure 19 describes the final ($T$ = 13 Gyr) locations of the four models 
on two color diagrams, ($V-K$)--($U-V$) and  ($I-K$)--($U-V$),
both for GISSEL96 and for KA SSP models. 
It is clear from this figure that our elliptical models can be located
in regions where the observed (Coma) E/S0s are located in the two color diagrams.
This result implies that if we adopt parameters suitable for
E/S0 formation (i.e., very small $T_{\rm trun}$ and ${\tau}_{\rm in}$)
our chemophotometric evolution model
can reproduce the observed global colors of the present day E/S0s 
and thus that the present model is rather realistic and reasonable
in describing both late- and early-type galaxies.


\clearpage

\begin{figure}
\epsscale{0.7}
\plotone{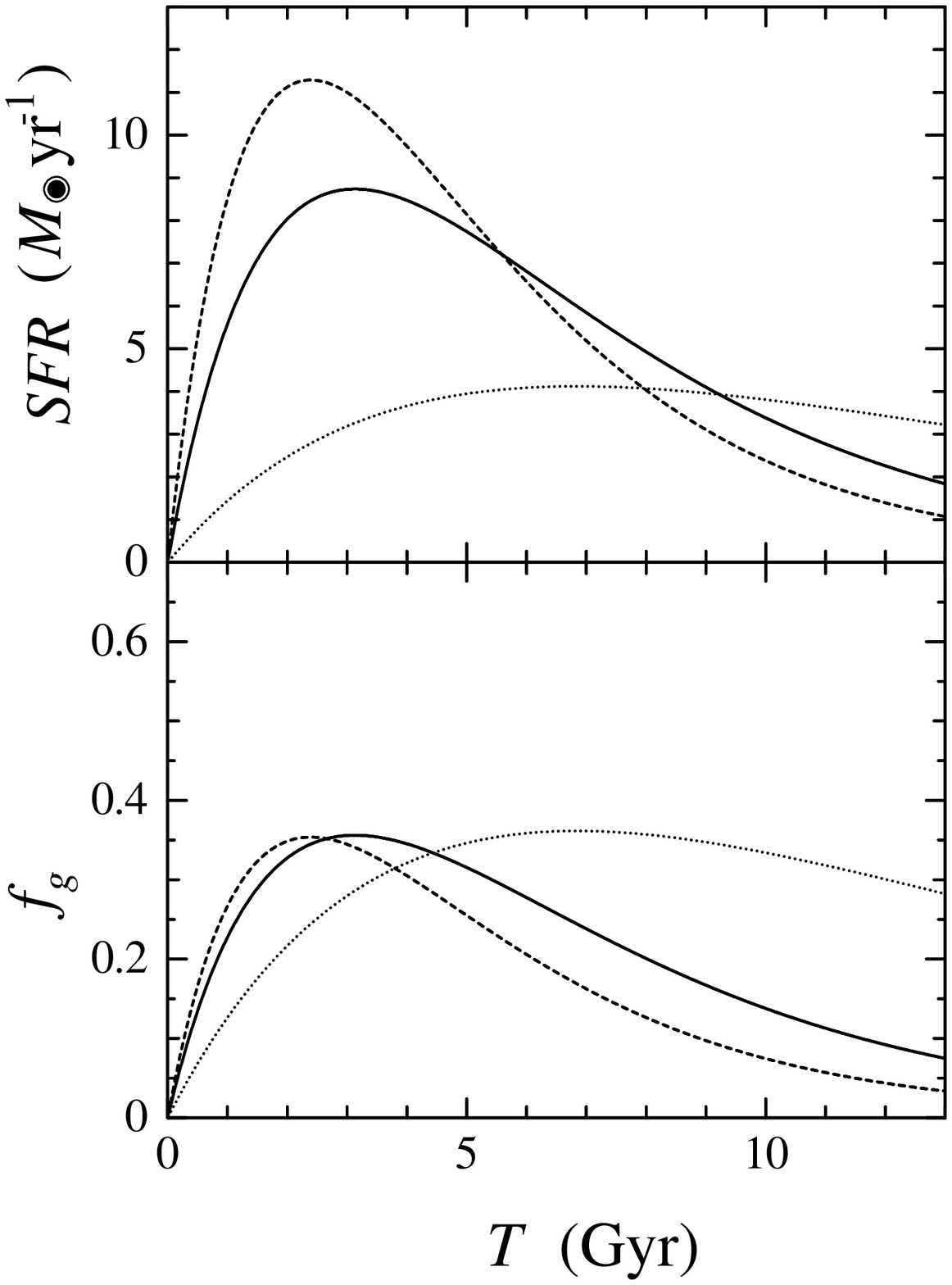}
\caption{
The star formation history (upper panel)
and the time evolution of gas mass fraction (lower one) 
for the three disk models, Sa (dashed), Sb (solid), and Sc (dotted).
Here the CF (continuously star-forming) models are plotted.
The mass of the model galaxy is assumed to be 
$6 \times 10^{10}M_{\odot}$ (corresponding to the
disk mass of the Galaxy).
}
\end{figure}

\begin{figure}
\epsscale{0.7}
\plotone{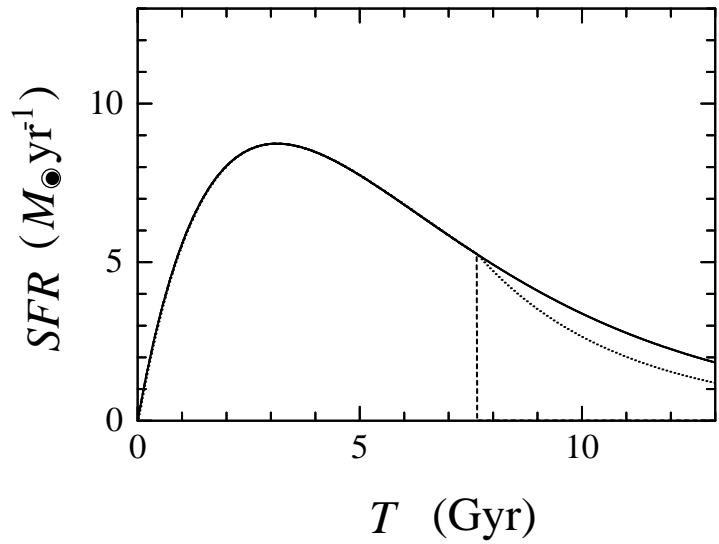}
\caption{
Star formation histories of the fiducial Sb models with 
different epochs of truncation of star formation: TF model, represented
by the {\it dashed} line), TI model represented by the {\it dotted} line.
For comparison, the Sb model with no truncation (CF model, {\it solid}
line) is superimposed.
} 
\end{figure}

\begin{figure}
\epsscale{0.7}
\plotone{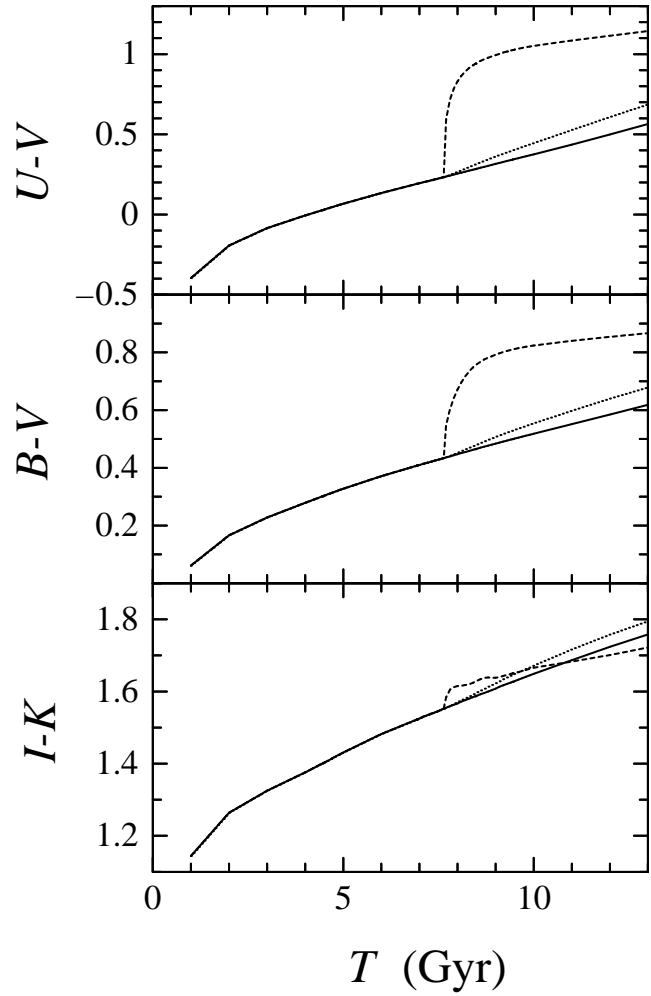}
\caption{
Color evolution in $U-V$ (top), $B-V$ (middle), and $I-K$ (bottom)
for the  three Sb disk models. Here the CF (continuously star-forming), 
TF (truncation of star formation), TI (truncation of gas infall) models
are represented by solid, dashed, and dotted lines, respectively. 
} 
\end{figure}

\begin{figure}
\epsscale{0.7}
\plotone{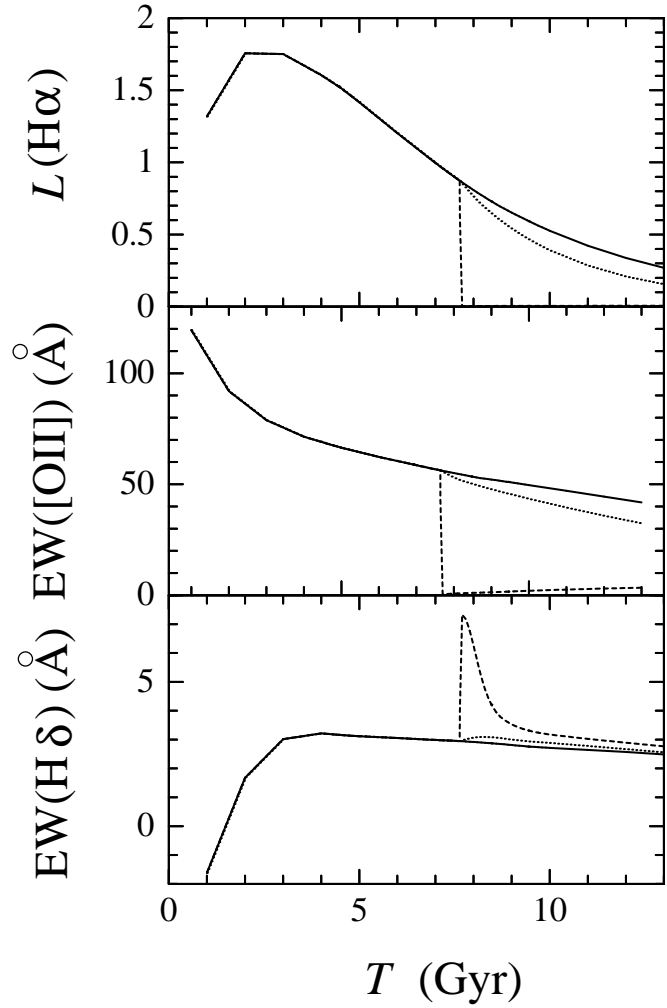}
\caption{
Time  evolution of H$\alpha$ luminosity (top, in units of $10^{35}$W),
EW([OII]) (middle, \AA), and EW(H$\delta$) (bottom, \AA) 
for the CF, TF, and TI models in the fiducial Sb-type star formation history. 
} 
\end{figure}

\begin{figure}
\epsscale{0.7}
\plotone{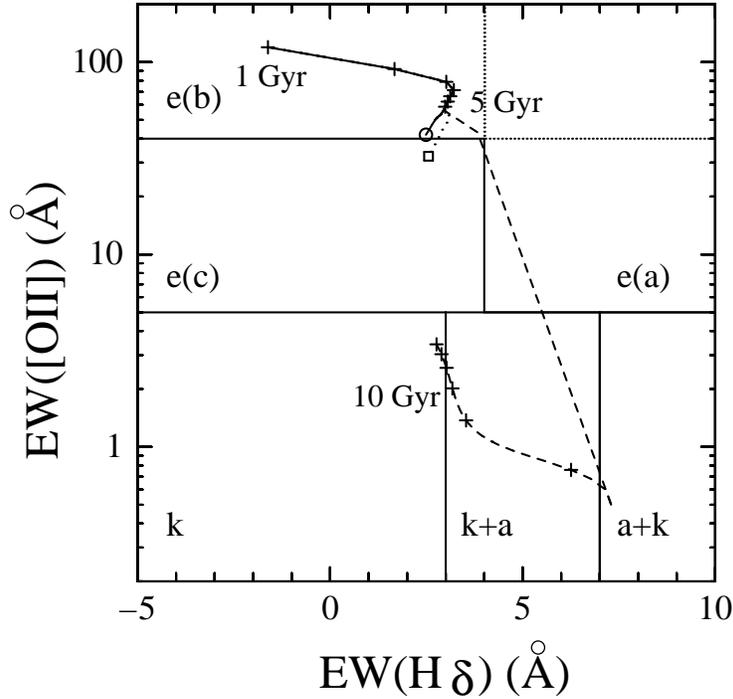}
\caption{
The evolution of galaxies on the EW([OII]) - EW(H$\delta$) plane 
for the fiducial CF (solid), the TF (dashed), and the TI models  (dotted).
Here an Sb-type star formation history is assumed. 
A time sequence, with intervals of 1\,Gyr, is indicated by {\it plus} 
signs along each locus of the TF model. 
In order to show more clearly the detailed evolution for 7.64 $\le$ $T$
$\le$ 7.65 Gyr, the time sequence with the interval of 0.001 Gyr is indicated
by {\it crosses} for the TF model.
The boundaries which define the different spectral types of Dressler et
al. (1999) are also superimposed. All of the models show e(b) spectra initially.
Note that the spectral type evolves
from e(b), to e(a), to a+k, to k+a, and finally
to k and also that the time scale for the TF model to show an e(a) spectrum
is rather short (less than 0.004 Gyr).
} 
\end{figure}

\begin{figure}
\epsscale{0.7}
\plotone{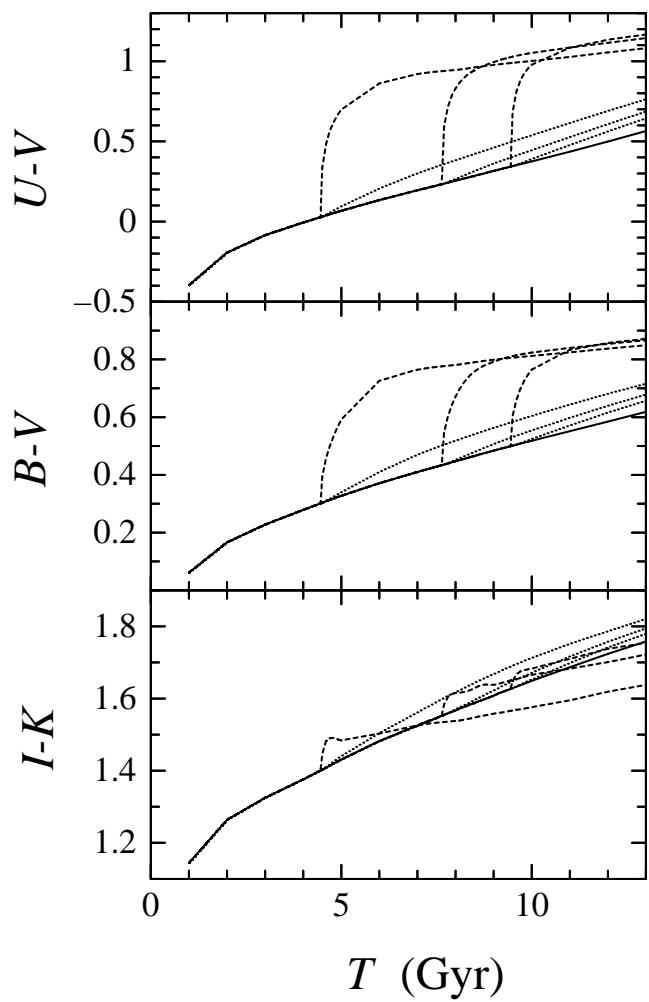}
\caption{
Dependence of 
the color evolution in $U-V$ (top), $B-V$ (middle), and $I-K$ (bottom)
on the epoch of truncation
for the CF (solid), the TF (dashed), and the TI (dotted) models in the case of
Sb-type star formation. Here the results of the models
with $T_{\rm trun}$ = 9.45, 7.64, and 4.46 Gyr
($z_{\rm trun}$ = 0.2, 0.4, and 1.0) for the TF and the TI models are given. 
} 
\end{figure}

\begin{figure}
\epsscale{0.7}
\plotone{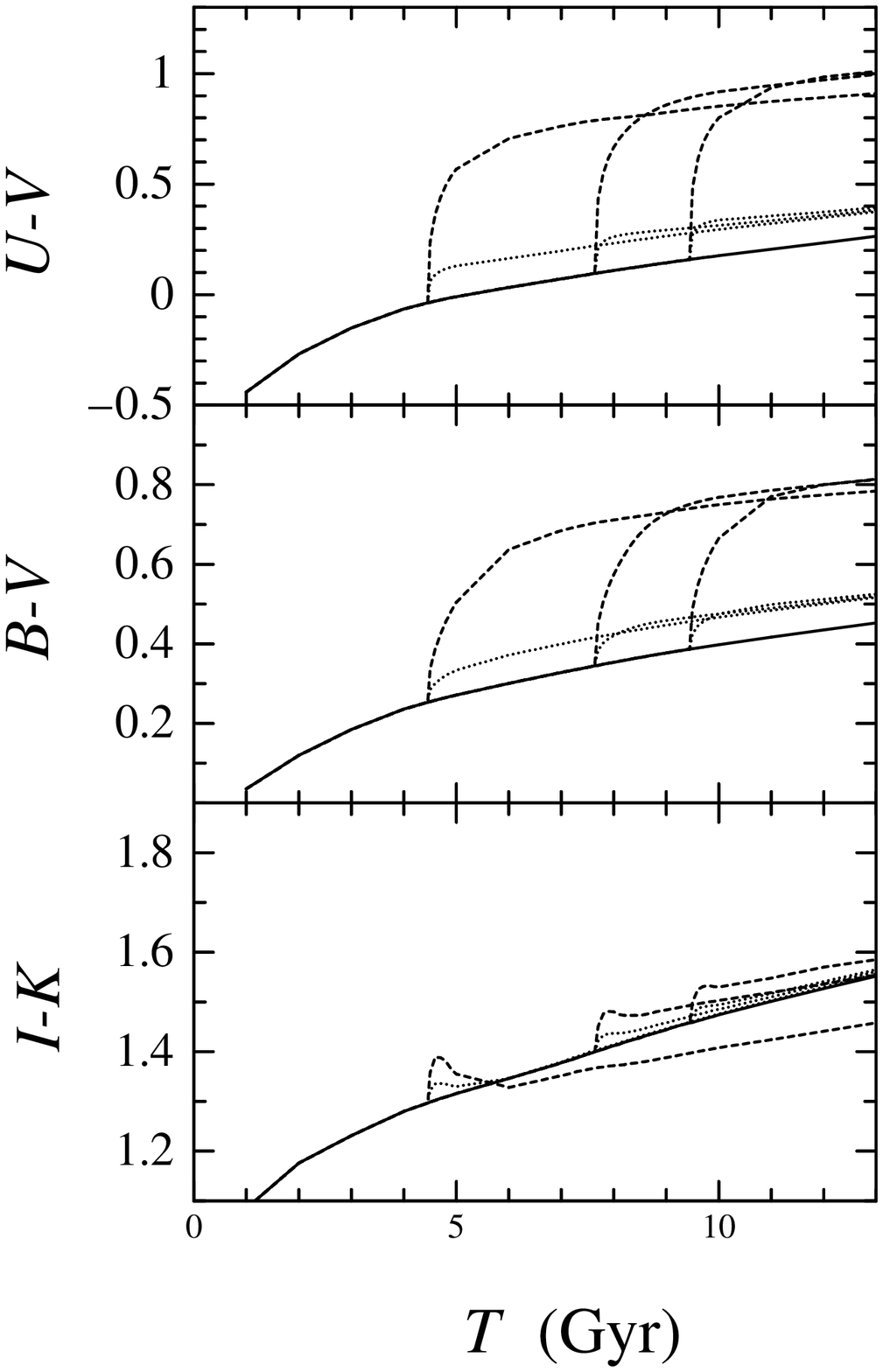}
\caption{
The same as Figure 6 but for the Sc models.
} 
\end{figure}

\begin{figure}
\epsscale{0.7}
\plotone{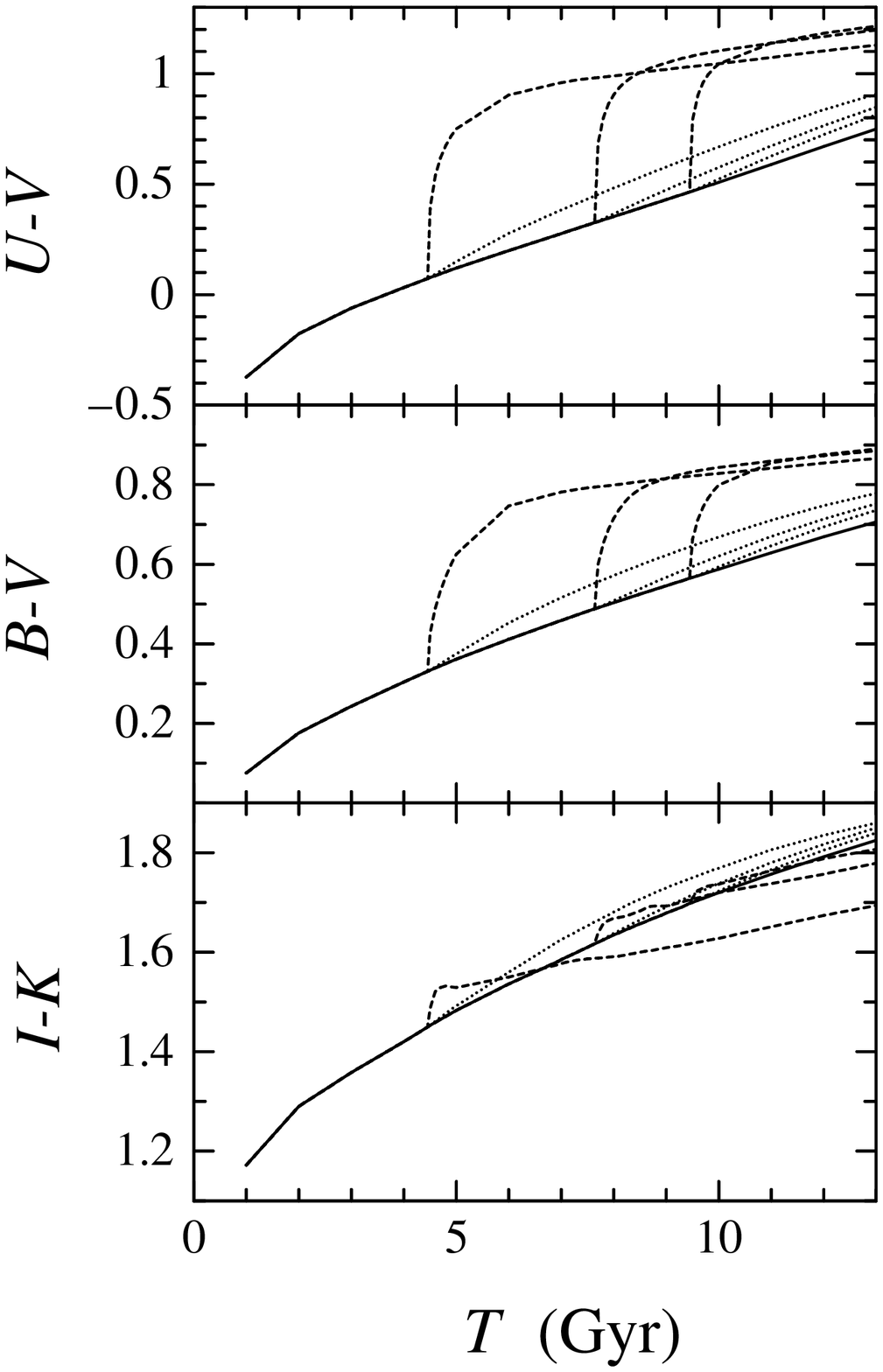}
\caption{
The same as Figure 6 but for the Sa models.
} 
\end{figure}

\begin{figure}
\epsscale{0.7}
\plotone{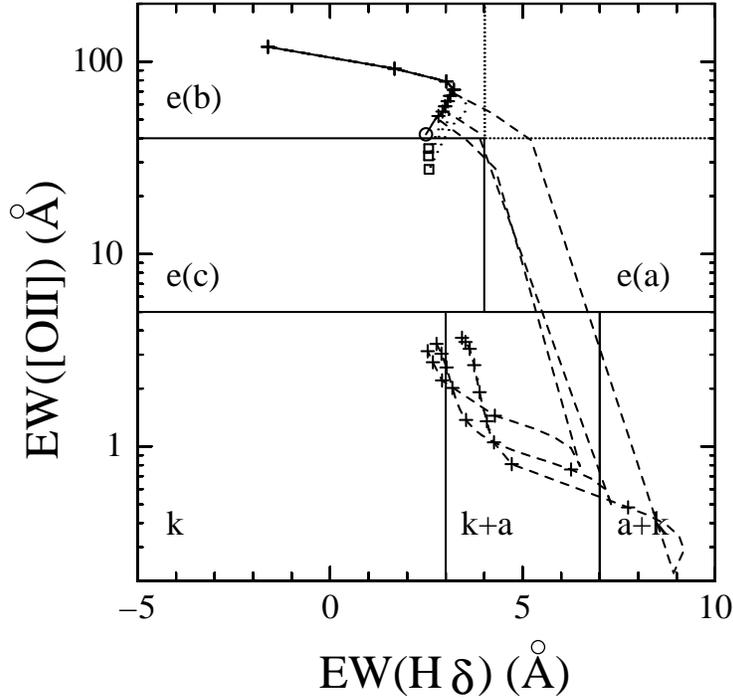}
\caption{
The dependence of 
the evolution  on the EW([OII]) - EW(H$\delta$) plane 
on the epoch of truncation represented by $z_{\rm trun}$
for the fiducial CF (solid), the TF (dashed), and the TI model (dotted).
Here the results of the models
with $T_{\rm trun}$ = 9.45, 7.64, and 4.46 Gyr
($z_{\rm trun}$ = 0.2, 0.4, and 1.0)
are given and Sb-type star formation history is assumed. 
Time sequence with the interval of 1 Gyr is indicated by plus
along the lines of the TF model.
The final results at $T$ = 13 Gyr  for CF and TI models are represented
by an open circle and an open square, respectively. 
The criteria of classification by Dressler et al. (1999) 
are also superimposed. All of the models show e(b) spectra initially.
} 
\end{figure}

\begin{figure}
\epsscale{0.7}
\plotone{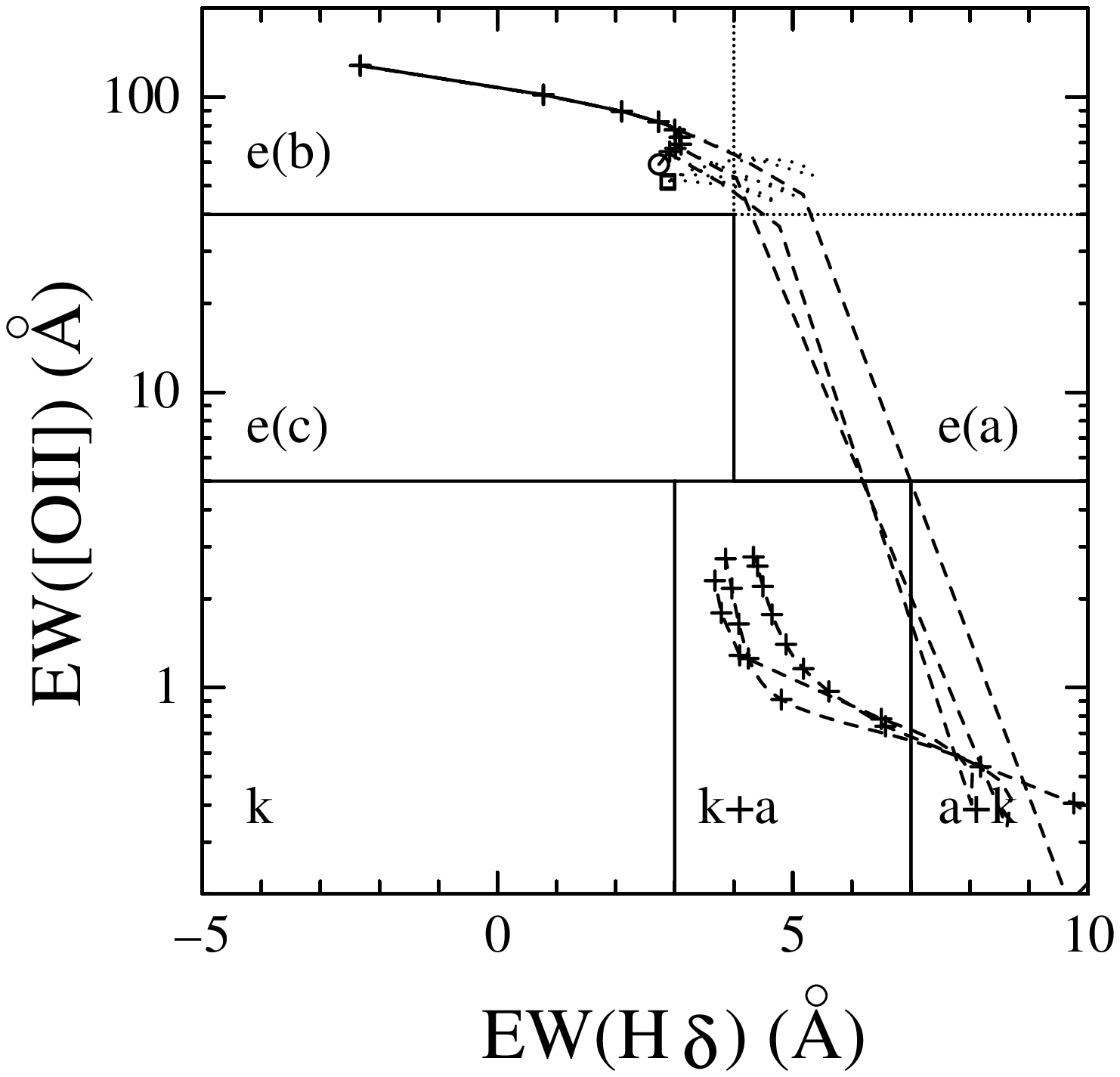}
\caption{
The same as Figure 9 but for the Sc models.
} 
\end{figure}

\begin{figure}
\epsscale{0.7}
\plotone{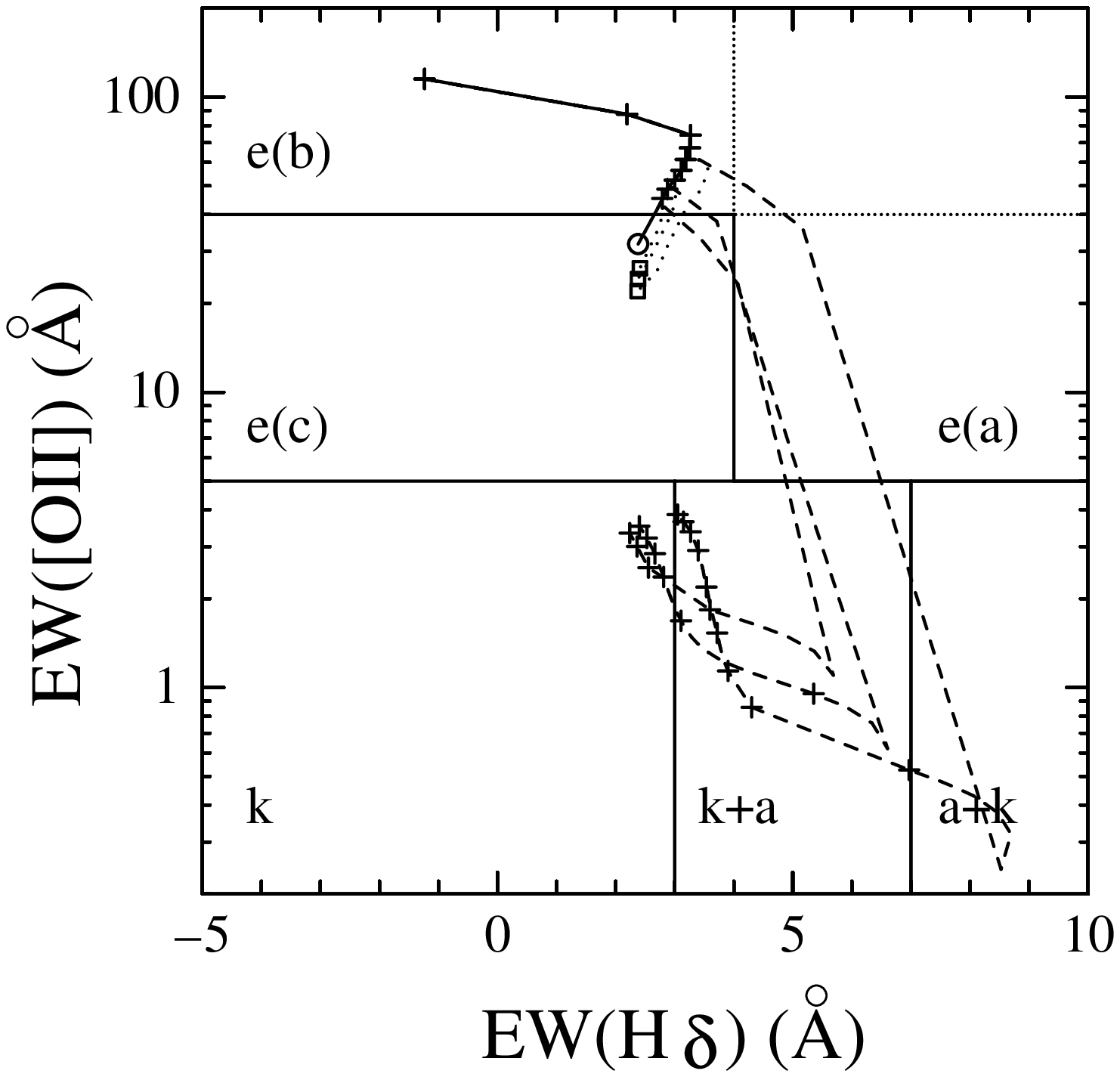}
\caption{
The same as Figure 9 but for the Sa models.
} 
\end{figure}

\begin{figure}
\epsscale{0.5}
\plotone{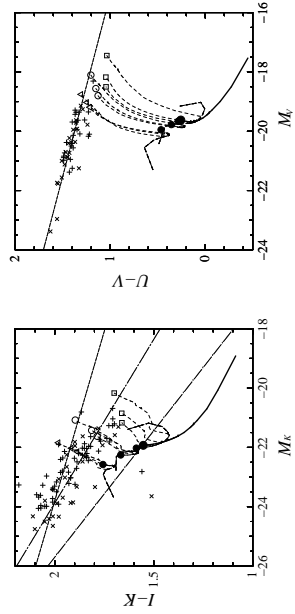}
\caption{
Comparison of the present numerical results 
with the observed ($I-K$)--$M_{\rm K}$ color-magnitude relation 
of the present-day E/S0s (left frame)
and the ($U-V$)--$M_{\rm V}$ one (right frame). 
Here, the observed  $I-K$ color-magnitude relation
relation in the Coma cluster by Eisenhardt et al. (2001; 
{\it short-long-dashed} line for S0 and {\it dotted} for E) and by de Grijs \& Peletier (1999; 
{\it short-dashed} line) are plotted  in the left  panel.
The observed $U-V$ color-magnitude relation by Bower et al. (1992)
is plotted in the left panel.
For both of these panels, observational data data points are also given 
separately for Es (cross) and for S0s (plus).
For comparison, the color-magnitude relation of disk galaxies
by de Grijs \& Peletier (1999) is also superimposed by a dot-long-dashed line.
Here the results of the TF model sequence S3 with 
Sc-type star formation 
and  variously different truncation epochs ($T_{\rm trun}$ or $z_{\rm trun}$)
and initial bulge mass fraction ($f_{\rm bul}$) are shown.
A larger filled circle represents the final color and magnitude at $z$ = 0.0
for the pure Sc disk model without truncation and without bulges
and the thick solid  line attached to
the filled circle tracks  the evolution
of the disk without truncation on the $I-K$ ($U-V$) color-magnitude plane.
The three smaller filled circles  represents the final color and magnitude at $z$ = 0.0
for the three Sc disk models with bulges ($f_{\rm bul}$ = 0.1, 0.3, 0.5)
and without truncation:
A filled circle plotted  in the redder and the brighter region in the color-magnitude
plane  corresponds to the model with a  bigger bulge. 
Long-dashed lines attached to the three smaller circles
track the evolution of 
 the three models without truncation on the color-magnitude plane
Three open triangles, circles, and squares represent the final results (colors and magnitudes)
for $f_{\rm bul}$ = 0.1, 0.3, and 0.5, respectively, 
for TF models with different $T_{\rm trun}$.  
A dashed line connected between  smaller filled circle and each  of three open 
triangles (circle and square)   
represents  the evolutionary truck of each of models with given $T_{\rm trun}$.
} 
\end{figure}

\begin{figure}
\epsscale{1.0}
\plotone{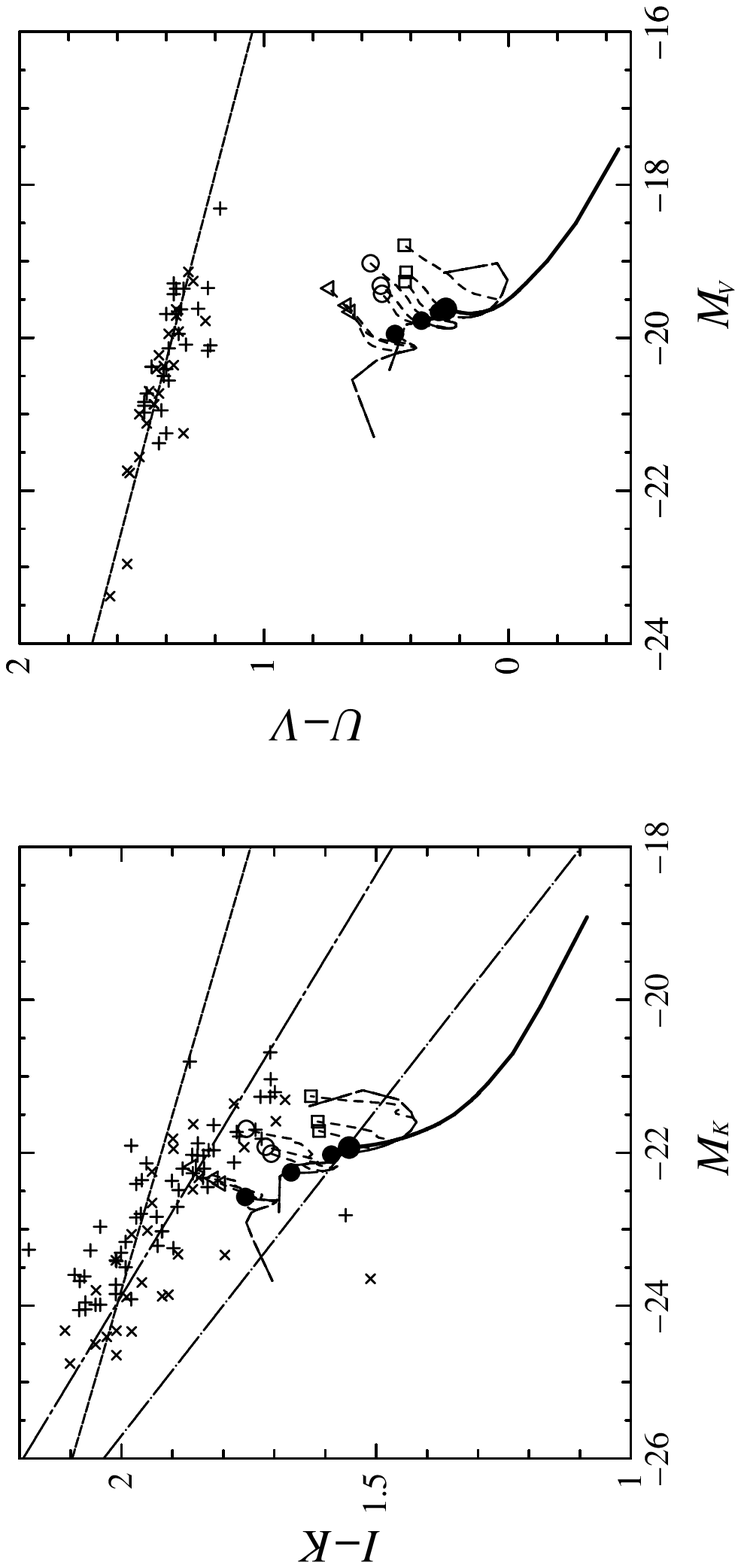}
\caption{
The same as Figure 12 but for the model sequence S3 with TI models.
} 
\end{figure}

\begin{figure}
\epsscale{1.0}
\plotone{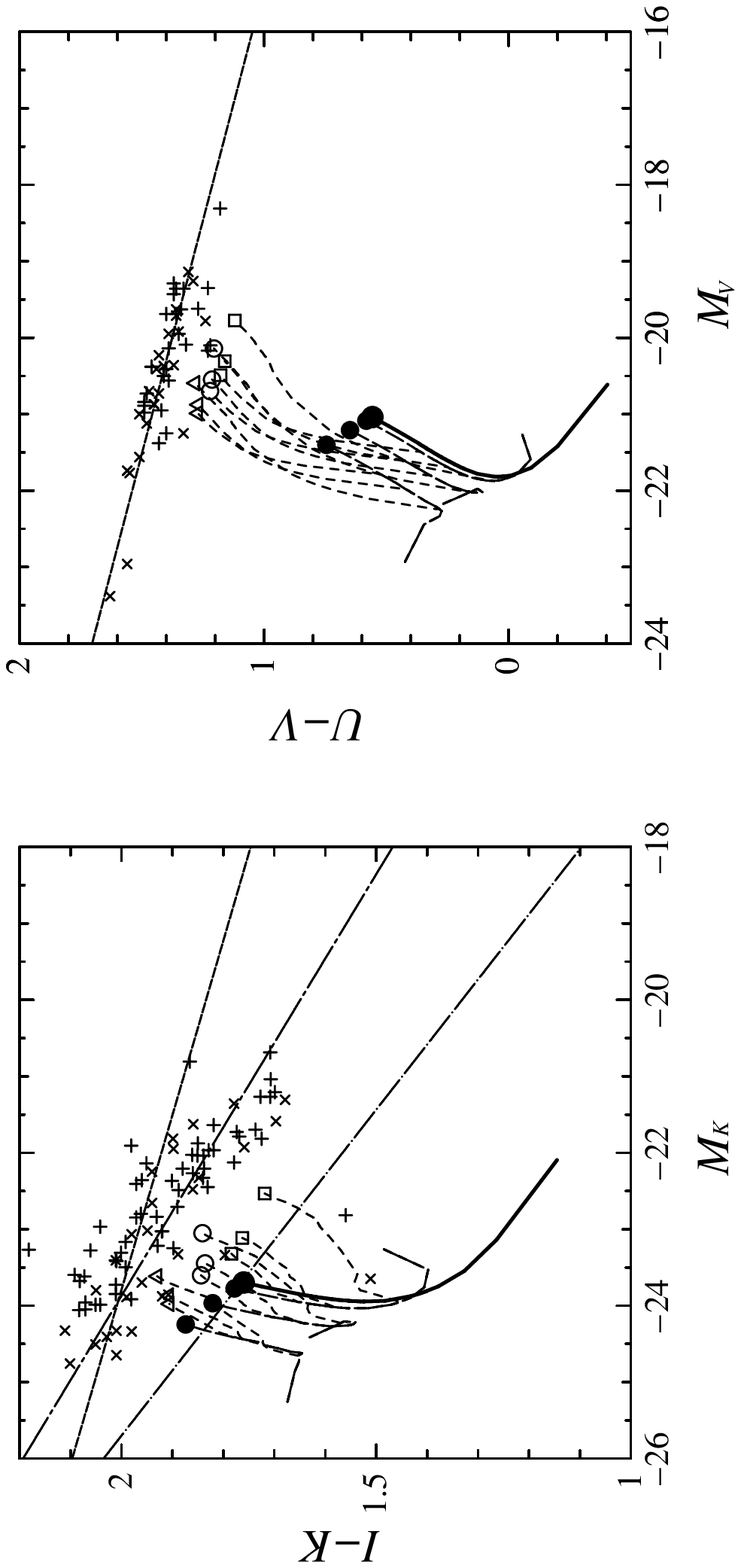}
\caption{
The same as Figure 12 but for the model sequence S2 with TF models.
} 
\end{figure}

\begin{figure}
\epsscale{1.0}
\plotone{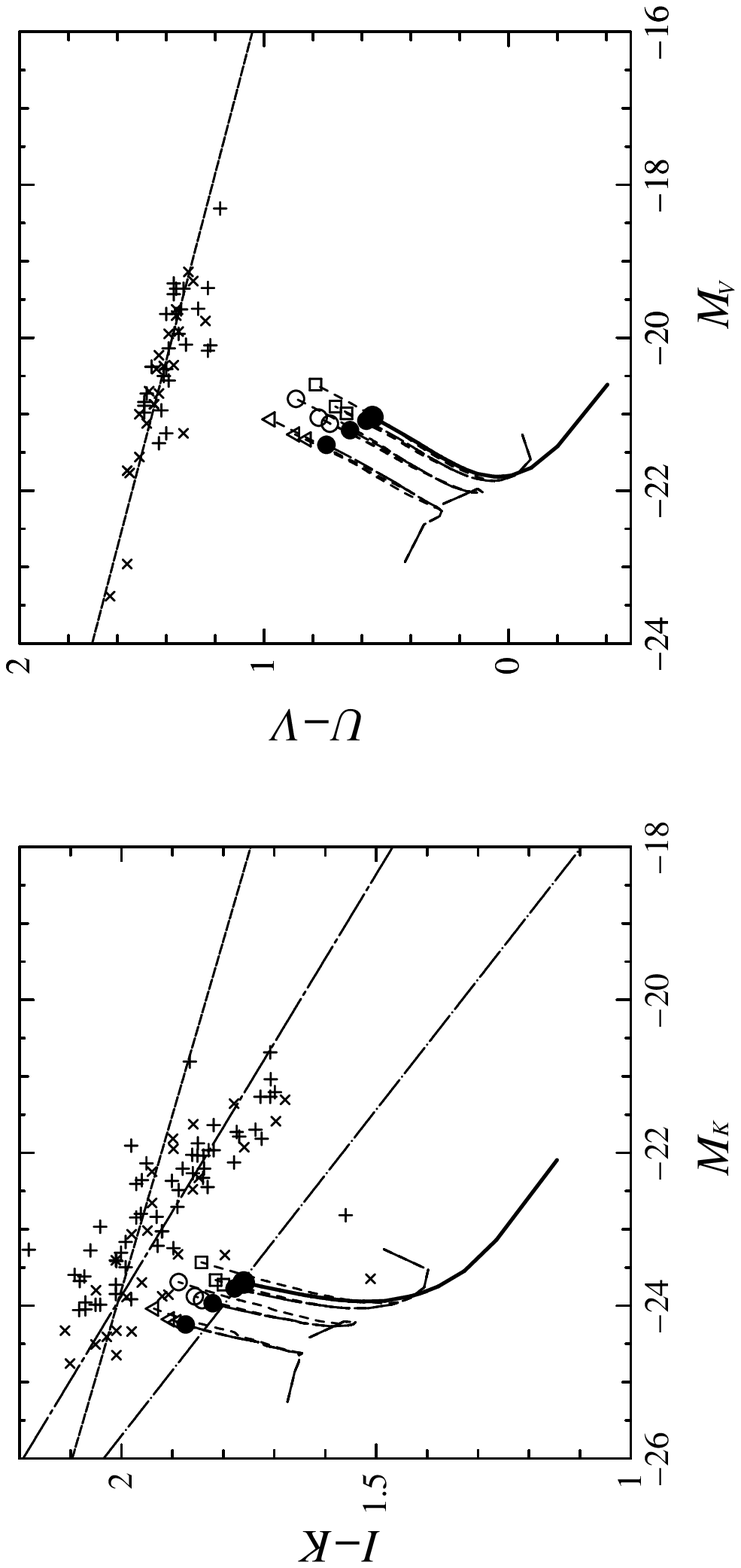}
\caption{
The same as Figure 12 but for the model sequence S2 with TI models.
} 
\end{figure}

\begin{figure}
\epsscale{1.0}
\plotone{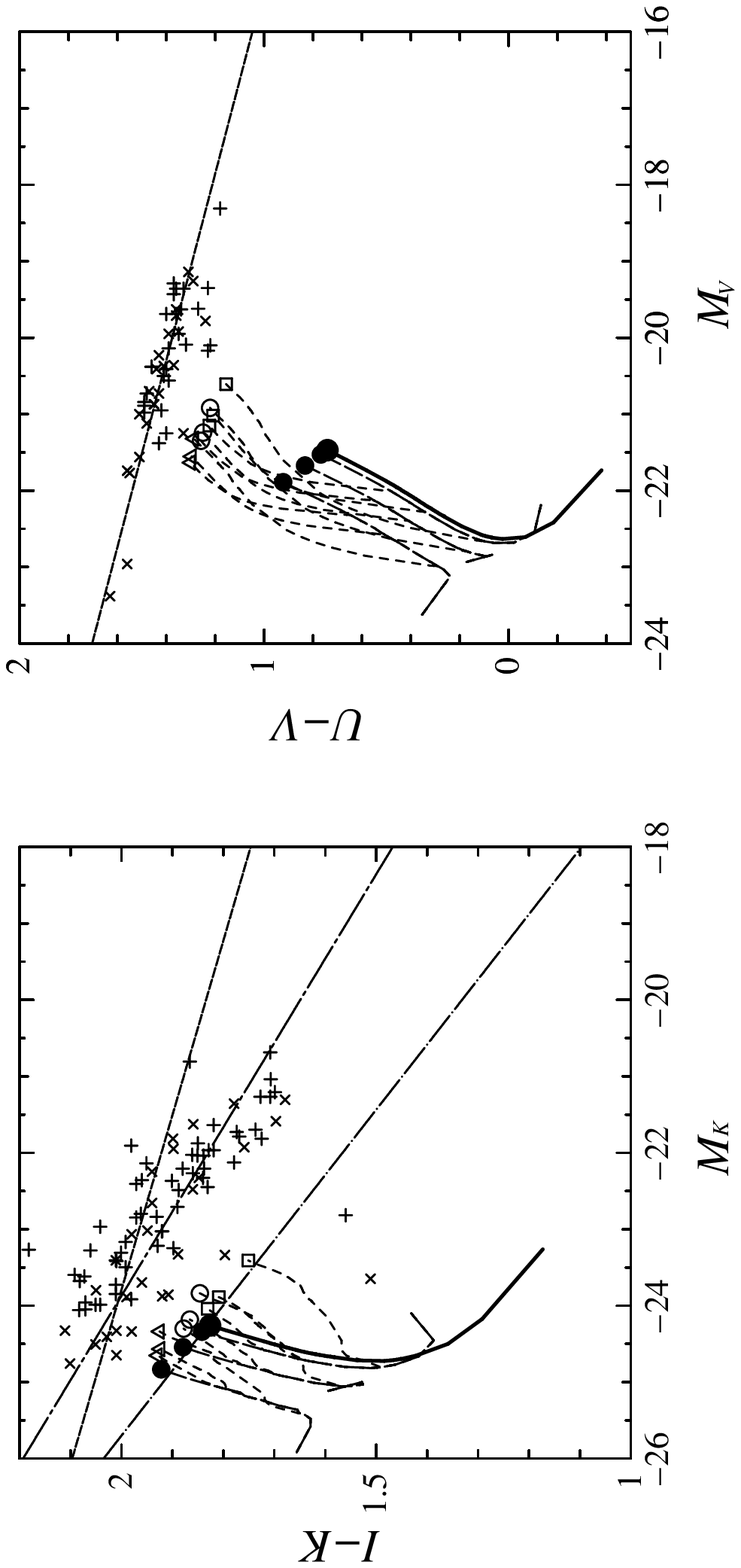}
\caption{
The same as Figure 12 but for the model sequence S1 with TF models.
} 
\end{figure}

\begin{figure}
\epsscale{1.0}
\plotone{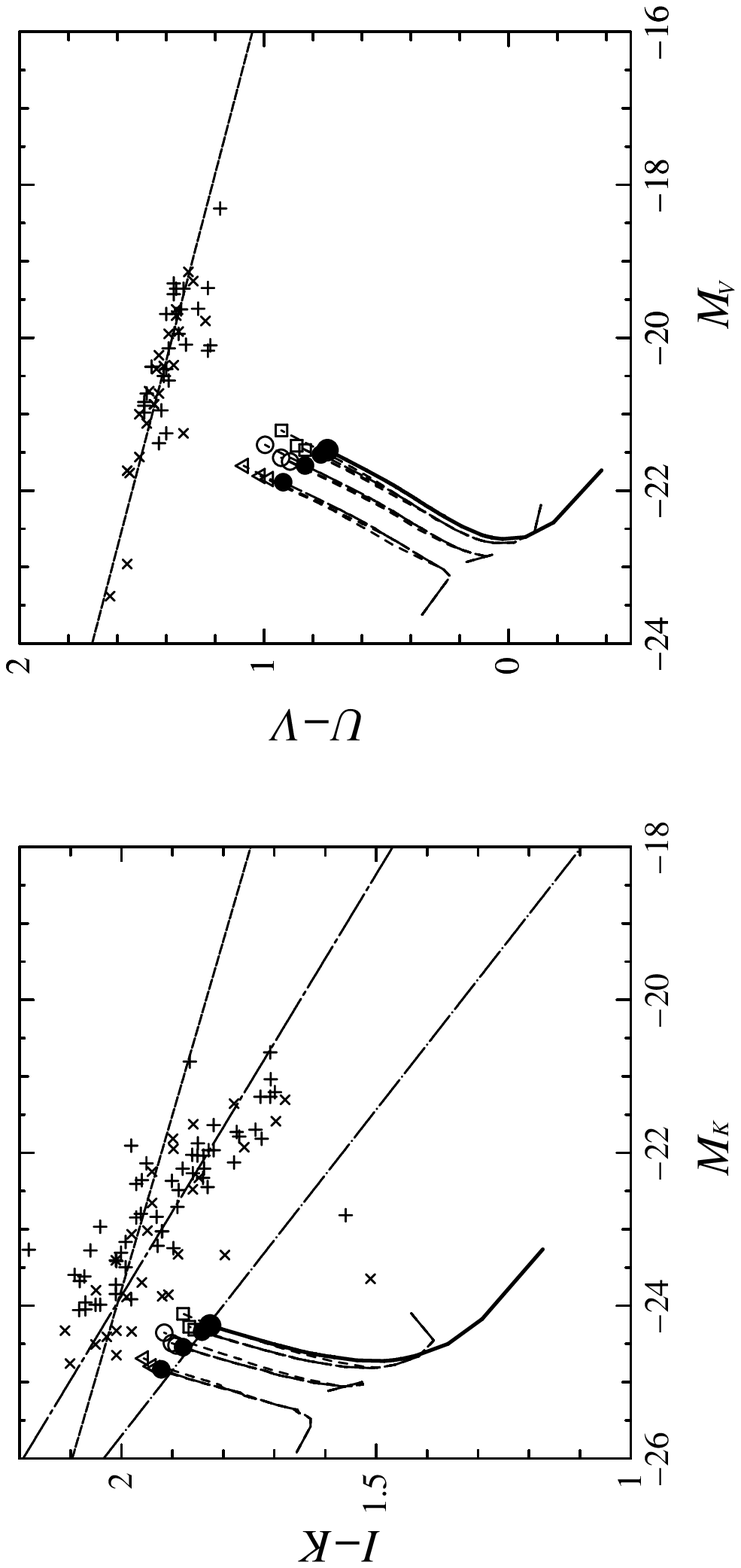}
\caption{
The same as Figure 12 but for the model sequence S1 with TI models.
} 
\end{figure}

\begin{figure}
\epsscale{0.7}
\plotone{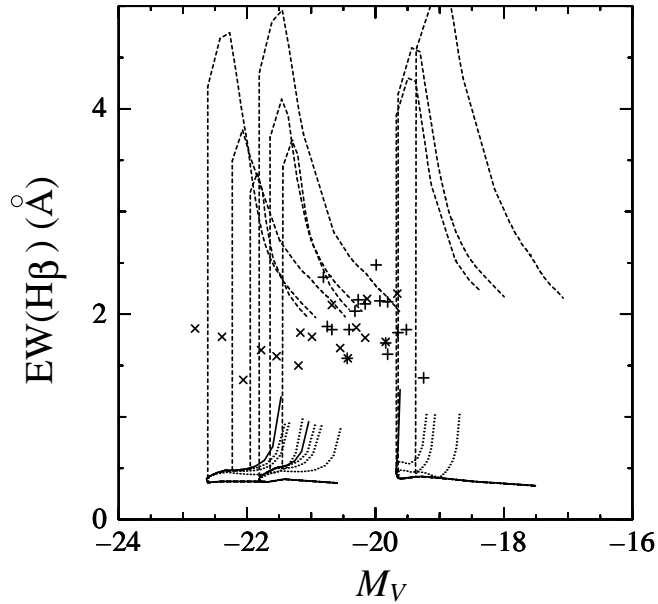}
\caption{
Evolution of truncated spiral models (TI  and TF with three
truncation epochs $T_{\rm trun}$ = 4.46, 7.64, and 9.45 Gyr) 
without bulges (i.e., $f_{\rm bul}$ = 0)
and continuously star-forming ones (CF) on $M_{\rm V}$-H${\beta}$ plane:
Solid, dotted, dashed lines describe CF, TI, and TF models, respectively,
for each star formation
history (i.e., Sa, Sb, and Sc models). 
Here brighter models corresponds to those with earlier type star formation
and models with later truncation show brighter $M_{\rm V}$.
For comparison, observational data by J{\o}rgensen (1999) are also
plotted by filled circles for Es and  for open ones for S0s.
} 
\end{figure}

\clearpage

\begin{figure}
\epsscale{0.7}
\plotone{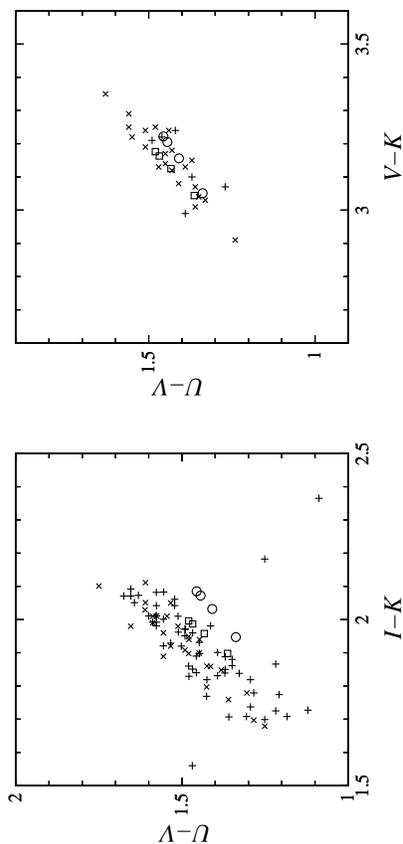}
\caption{
Locations of our elliptical galaxy models 
on the two color diagrams ($V-K$)--($U-V$) (right) and  ($I-K$)--($U-V$) (left).
The results based on SSPs by GISSEL96 and on those by Kodama \& Arimoto (1997)
are represented by circles and squares, respectively.
For comparison, observational results of Coma cluster E/S0s 
by Bower et al. (1992) and Eisenhardt et al. (2001) are plotted by crosses for
Es and by pluses for S0s.  
} 
\end{figure}

\clearpage

\begin{deluxetable}{ccc}
\footnotesize
\tablecaption{Set of models  for S0 color-magnitude relations \label{tbl-1}}
\tablewidth{0pt}
\tablehead{
\colhead{Model sequence} & 
model & \colhead{bulge  fraction $f_{\rm bul}$}}
\startdata
S1 & 
${\rm Sa}_{\rm TI}^{4.46}$, ${\rm Sa}_{\rm TI}^{7.64}$, ${\rm Sa}_{\rm TI}^{9.45}$, 
${\rm Sa}_{\rm TF}^{4.46}$, ${\rm Sa}_{\rm TF}^{7.64}$, ${\rm Sa}_{\rm TF}^{9.45}$, 
${\rm Sa}_{\rm CF}$ 
& 0.1, 0.3, 0.5\\
S2 & 
${\rm Sb}_{\rm TI}^{4.46}$, ${\rm Sb}_{\rm TI}^{7.64}$, ${\rm Sb}_{\rm TI}^{9.45}$, 
${\rm Sb}_{\rm TF}^{4.46}$, ${\rm Sb}_{\rm TF}^{7.64}$, ${\rm Sb}_{\rm TF}^{9.45}$, 
${\rm Sb}_{\rm CF}$ 
& 0.1, 0.3, 0.5\\
S3 & 
${\rm Sc}_{\rm TI}^{4.46}$, ${\rm Sc}_{\rm TI}^{7.64}$, ${\rm Sc}_{\rm TI}^{9.45}$, 
${\rm Sc}_{\rm TF}^{4.46}$, ${\rm Sc}_{\rm TF}^{7.64}$, ${\rm Sc}_{\rm TF}^{9.45}$, 
${\rm Sc}_{\rm CF}$ 
& 0.1, 0.3, 0.5\\
\enddata
\end{deluxetable}

\end{document}